\documentclass[sigconf,screen]{acmart}

\setcopyright{acmlicensed}
\acmPrice{15.00}
\acmDOI{10.1145/3468264.3468554}
\acmYear{2021}
\copyrightyear{2021}
\acmSubmissionID{fse21main-p146-p}
\acmISBN{978-1-4503-8562-6/21/08}
\acmConference[ESEC/FSE '21]{Proceedings of the 29th ACM Joint European Software Engineering Conference and Symposium on the Foundations of Software Engineering}{August 23--28, 2021}{Athens, Greece}
\acmBooktitle{Proceedings of the 29th ACM Joint European Software Engineering Conference and Symposium on the Foundations of Software Engineering (ESEC/FSE '21), August 23--28, 2021, Athens, Greece}

\usepackage[]{algorithm2e}
\SetKwInput{KwData}{Input}
\SetKwInput{KwResult}{Output}

\usepackage{multirow}
\usepackage{multicol}
\usepackage{amsmath}
\usepackage{pifont}
\usepackage{tabularx}
\usepackage{ragged2e}
\usepackage{booktabs,caption}
\usepackage[flushleft]{threeparttable}
\DeclareMathOperator*{\argmin}{\arg\min}
\usepackage{enumitem}

\newcommand\expName[1]{exploration tarpit}
\newcommand\expNames[1]{exploration tarpits}
\newcommand\ExpName[1]{Exploration tarpit}
\newcommand\ExpNames[1]{Exploration tarpits}
\newcommand\Ineffective[1]{Ineffective}
\newcommand\ineffective[1]{ineffective}
\newcommand\ExpSpacePartition[1]{Exploration Space Partition}
\newcommand\TrappedLocal[1]{Excessive Local Exploration}
\newcommand\numofapps[1]{16}
\newcommand\numofcombs[1]{48}
\newcommand\numoftraces[1]{144}
\newcommand\numofcombswithissue[1]{96}
\newcommand\toller[1]{\textsc{Toller}}
\newcommand\ourtool[1]{\textsc{Vet}}
\newcommand\avgCovImproveMax[1]{15.3\%}
\newcommand\avgCovPPIncreaseMax[1]{20.8}
\newcommand\cumCovImproveMax[1]{15.3\%}
\newcommand\cumCovPPIncreaseMax[1]{15.3}
\newcommand\allCrashImproveMax[1]{2.1x}
\newcommand\maxRegionLength[1]{98.6\%}
\newcommand\covImprApe{4.4\%}
\newcommand\covImprMonkey{15.3\%}
\newcommand\covImprWC{11.3\%}
\newcommand\crashImprApe{2.1x}
\newcommand\crashImprMonkey{2.1x}
\newcommand\crashImprWC{1.9x}
\newcommand\numOfRegions[1]{131}
\newcommand\avgRegionLengthMinutes[1]{about 27}

\newcommand{\cmark}{\ding{51}}%
\newcommand{\xmark}{\ding{55}}%

\begin{document}

\title{\ourtool{}: Identifying and Avoiding UI Exploration Tarpits}

\author{Wenyu Wang}
\affiliation{University of Illinois at Urbana-Champaign \country{USA}}
\email{wenyu2@illinois.edu}

\author{Wei Yang}
\affiliation{University of Texas at Dallas \country{USA}}
\email{wei.yang@utdallas.edu}

\author{Tianyin Xu}
\affiliation{University of Illinois at Urbana-Champaign \country{USA}}
\email{tyxu@illinois.edu}

\author{Tao Xie}\authornote{Tao Xie is with the Key Laboratory of High Confidence Software Technologies (Peking University), Ministry of Education, China, and is the corresponding author.}
\affiliation{Peking University \country{China}}
\email{taoxie@pku.edu.cn}

\begin{abstract}
Despite over a decade of research, it is still challenging for mobile UI testing
  tools to achieve satisfactory effectiveness, especially on industrial apps
  with rich features and large code bases.
Our experiences suggest that existing mobile UI testing tools are prone to \textit{\expNames{}}, where the tools get stuck with a small fraction of app functionalities for an extensive amount of time.
For example, a tool logs out an app at early stages without being able to log back in, and since then the tool gets stuck with exploring the app's pre-login functionalities (i.e., \expNames{}) instead of its main functionalities. 
While tool vendors/users can manually hardcode rules for the tools to avoid specific \expNames{}, these rules can hardly generalize, being fragile in face of diverted testing environments, fast app iterations, and the demand of batch testing product lines.
To identify and resolve \expNames{}, we propose \ourtool{}, a general approach including a supporting system for the given specific Android UI testing tool on the given specific app under test (AUT). 
\ourtool{} runs the tool on the AUT for some time and records UI traces, based on which \ourtool{} identifies \expNames{} by recognizing their patterns in the UI traces.
\ourtool{} then pinpoints the actions (e.g., clicking logout) or the screens that lead to or exhibit \expNames{}.
In subsequent test runs, \ourtool{} guides the testing tool to prevent or recover from \expNames{}.
From our evaluation with state-of-the-art Android UI testing tools on popular industrial apps, \ourtool{} identifies \expNames{} that cost up to \maxRegionLength{} testing time budget.
These \expNames{} reveal not only limitations in UI exploration
  strategies but also defects in tool implementations. %
\ourtool{} automatically addresses the identified \expNames{}, enabling each evaluated tool to achieve higher code coverage and improve crash-triggering capabilities.
\end{abstract}



\maketitle

\section{Introduction}
\label{sec:intro}

With the prosperity of mobile apps~\cite{android_market_share},
  especially their roles in people's daily life during pandemic
  (e.g., food ordering, grocery delivery, and social networking),
  quality assurance of mobile apps becomes crucially important.
User Interfaces (UIs), as the primary interface of user-app interactions, are natural entry points for app testing.
While manual UI testing is still often used in practice, automated UI testing is becoming popular~\cite{ten_best_tools,how_to_perform_mobile_automation_testing,How-to-Automate-Mobile-App-Testing,mastering_the_art_of_mobile_testing}.
Automated UI testing mimics how human users interact with apps through the UIs and detects reliability and usability issues.
Automated UI testing complements manual testing with greater timing flexibility and better code coverage, requiring little human intervention.

However, existing mobile UI testing tools are found to be ineffective in exploring app functionalities, despite their sophisticated strategies for UI exploration.
While recent proposals~\cite{Mao:2016,su2017guided,Choi:2013,Hao:2014,yang13:grey,Machiry:2013,Ye:2013:DFA:2536853.2536881,DroidBot,Azim:2013,EvoDroid,Anand:2012} have reported promising results,
  measurement studies on comprehensive app benchmarks~\cite{ChoudharyGO15,Wang:2018:ESA:3238147.3240465} have drawn different conclusions.
For example, a recent study~\cite{Wang:2018:ESA:3238147.3240465} shows that state-of-the-art mobile UI testing tools yield low code coverage (about 30\% in method coverage) after hours of testing on popular industrial apps.
Note that industrial apps typically have richer functionalities and larger code bases, compared with open-source apps.
The findings suggest a significant effectiveness gap that needs to be filled for automated mobile UI testing.

According to our experience, the ineffectiveness of existing mobile UI testing tools often stems from their proneness to \textit{\expNames{}}\footnote{The name of \expNames{} is inspired by the Mythical Man-Month book~\cite{brooks95:mythical}.}, where tools get stuck with a small fraction of app functionalities for an extensive amount of time.
We show a real-world example in \S\ref{sec:example}, where a
    state-of-the-art Android UI testing tool named Ape~\cite{Gu:2019:PGT:3339505.3339542} decides to log itself out one minute after testing an app starts, without being able to log back in, and since then gets stuck with exploring the app's pre-login functionalities (i.e., \expNames{}) instead of its main functionalities. 
It is possible that tool vendors/users manually hardcode rules for the tools to avoid specific \expNames{}, such as instructing Ape to avoid tapping the ``logout'' button or writing a script to support automatic login.
However, these rules can hardly generalize, being fragile in face of diverted testing environments (e.g., unreliable network to process login requests), fast app iterations, and the demand of batch testing product lines.
Our findings in \S\ref{sub:root_causes_of_ineffective_exploration_intervals}
  show various cases as such where \expNames{} can be caused by unexpected flaws in a tool's exploration strategies and/or  implementation defects.

\begin{figure*}
\centering
\includegraphics[width=2\columnwidth]{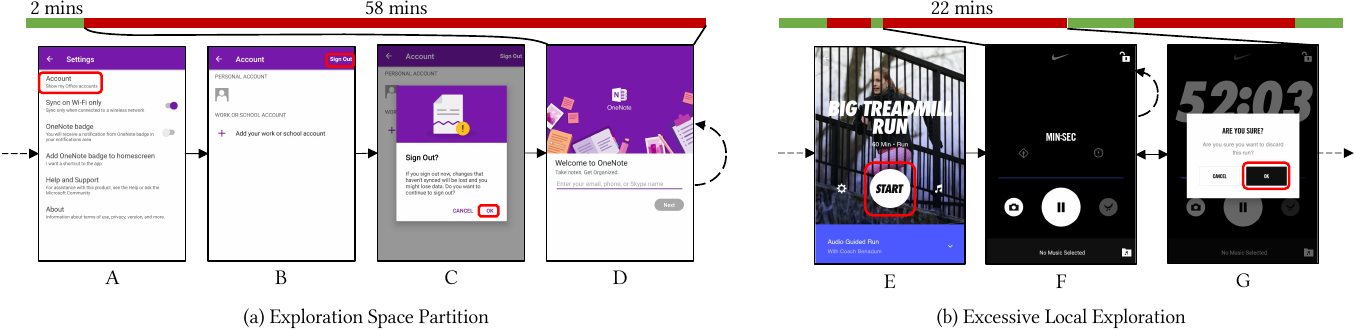}
\begin{justify}
\small{
Note:
Colored bars on the top represent the progress of two 1-hour tests, where green bars refer to normal exploration and red bars refer to \expNames{}.
Dashed straight arrows indicate visiting the screen from some other screen, and solid arrows show transitions between two screens after clicking the red-boxed UI elements.
The dashed curve arrow on Screen D depicts that Ape cycles around D until the end of testing.
The dashed curve arrow on Screen F shows that Monkey stays on F within the 22-minute \expName{} window.
}
\end{justify}
\caption{Motivating examples of \expNames{} described in \S\ref{sec:example}.
}
\Description[Motivating examples of \expNames{}.]
{Two motivation examples of \expNames{}, depicting \ExpSpacePartition{} and \TrappedLocal{}, respectively.}
\label{fig:ape_onenote_1}
\end{figure*}

To automatically identify and resolve
  \expNames{}, in this paper, we propose a general approach and its supporting system named \ourtool{} for the given specific Android UI testing tool on the given specific app under test (AUT). 
\ourtool{} works in three stages. (1) \ourtool{} runs the tool on the AUT for some time
  and records the interactions between the tool and AUT, in the form of UI \textit{traces}.
A UI trace consists of app UIs interleaving with the actions taken by the tool.
(2) \ourtool{} then analyzes the collected traces to identify trace subsequences (termed {\it regions})
  that manifest \expNames{}.
(3) \ourtool{} guides the tool in subsequent test runs to prevent or recover from an \expName{}
  by monitoring the testing progress and taking actions based on findings from the identified regions.

\ourtool{} includes two specialized algorithms targeting two corresponding patterns of \expNames{}: \textit{\ExpSpacePartition{}} and \textit{\TrappedLocal{}} (see \S\ref{sec:example} and \S\ref{sec:algorithms}).
\ExpSpacePartition{}, corresponding to Figure \ref{fig:ape_onenote_1}a, indicates that the fraction of app functionalities explored by the tool is disconnected from most of the app functionalities after some specific action (e.g., tapping ``OK'' in Screen C).
Such situations can be prevented by disabling the aforementioned action.
\TrappedLocal{} indicates that the tool enters a hard-to-escape fraction of the app UIs and needs a significant amount of time to reach other functionalities, as demonstrated in Figure~\ref{fig:ape_onenote_1}b.
This issue can be addressed by either preventing the tool from entering (e.g., disabling ``START'' in Screen E in Figure~\ref{fig:ape_onenote_1}b) or assisting the tool to escape (e.g., restart the app upon observation of Screen F).
To design the two algorithms, we first construct fitness value formulas that quantify how well a region on the given trace matches a  targeted pattern.
We then apply {fitness value optimization} on the entire trace to determine the region that best fits our targeted patterns.

We evaluate \ourtool{} using three state-of-the-art/practice Android UI testing tools
  (Monkey~\cite{AndroidMonkey}, Ape~\cite{Gu:2019:PGT:3339505.3339542}, and WCTester~\cite{Zeng:2016,zheng17:automated})
  with \numofapps{} widely used industrial apps.
We collect \numoftraces{} traces by running each tool on each app three times for one hour each
  (\textit{original} runs).
\ourtool{} reports at least one \expName{} region in each (tool, app) pair, with \numOfRegions{} regions in total, each spanning \avgRegionLengthMinutes{} minutes on average.
The longest regions span over 59 minutes, about \maxRegionLength{} of the one-hour testing time budget.
After inspecting the \numOfRegions{} reported regions,
  we confirm the root causes of \numofcombswithissue{} regions,
  including both limitations of UI exploration strategies (e.g., early logouts)
  and defects in tool implementation (e.g., hanging), as shown in \S\ref{sub:root_causes_of_ineffective_exploration_intervals}.
We then perform six other one-hour runs for each (tool, app) pair:
  (1) three \textit{guided} runs using \ourtool{} to automatically avoid all the \expName{} regions identified in the original runs during testing on three runs,
  and (2) three \textit{comparison} runs not using \ourtool{}.

Based on the preceding evaluation setup, we compare the code coverage (of the given app) achieved by applying each tool with and without the assistance of \ourtool{} given the same time budget.
Specifically, we compare the combined code coverage and the numbers of distinct crashes for
  (1) original runs and guided runs,
  and (2) original runs and comparison runs.
The evaluation results show that on average a tool assisted by \ourtool{} achieves up to a \cumCovImproveMax{} relative code coverage increment and triggers up to \allCrashImproveMax{} distinct crashes than the tool without the assistance of  \ourtool{}.

In summary, this paper makes the following main contributions:
\begin{itemize}[leftmargin=*]
\item A new perspective of improving the given automated UI testing tool by
  automatically identifying and addressing \expNames{} for the given target AUT;
\item Algorithms for effective identification of two manifestation patterns
  of \expNames{};
\item A practical system~\cite{vet-github} that can be automatically applied to enhance any Android UI testing tool such as Monkey~\cite{AndroidMonkey}, Ape~\cite{Gu:2019:PGT:3339505.3339542}, and WCTester~\cite{Zeng:2016,zheng17:automated}, on any AUT;
\item Comprehensive evaluation of \ourtool{}, demonstrating that \ourtool{} reveals various issues related to tools  or app usability, and that \ourtool{} automatically resolves those issues, helping the tools achieve up to a \avgCovImproveMax{} relative code coverage increment and \allCrashImproveMax{} distinct crashes on \numofapps{} popular industrial apps.
\end{itemize}

\section{Motivating examples}
\label{sec:example}

We present two concrete examples from our experiments covering \ExpSpacePartition{} and \TrappedLocal{} (see \S\ref{sec:algorithms}).
These examples provide contexts for further discussion and help illustrate the motivations that drive the design of \ourtool{}.

\subsection{\ExpSpacePartition{}}

We run Ape~~\cite{Gu:2019:PGT:3339505.3339542}, a state-of-the-art Android UI testing tool to test a popular app, Microsoft \textit{OneNote}.
The result is illustrated in Figure~\ref{fig:ape_onenote_1}a.
We manually set up the account to
  log in to the app's main functionalities,
  and then start {Ape}. We run {Ape} without interruptions for one hour and check the test results afterward.

In the one-hour testing period, {Ape} explores only {12\% (9 out of 76)} activities. To understand the low testing effectiveness, we investigate the UI trace captured during testing and find the root cause to be \expNames{}:

\begin{enumerate}[leftmargin=*]
\item
Ape performs exploration around OneNote's main functionalities for about two minutes, covering 7 (out of 9) of all the activities covered in the entire one-hour test run.
We omit this phase in Figure \ref{fig:ape_onenote_1}a.
\item
About two minutes after testing starts, Ape arrives at the ``Settings'' screen (Screen A) and decides to click ``Account'' (the red-boxed UI element) for further exploration.
\item
Ape arrives at the ``Account'' screen (Screen B) and clicks
the ``Sign Out'' button.
The click pops up a window (Screen~C) asking for confirmation of getting logged out.
\item
Ape clicks ``CANCEL'' first, and then goes back to the ``Account'' screen.
However, Ape clicks the ``Sign Out'' button again, knowing that there is one action not triggered yet in the confirmation dialog.
Subsequently, Ape clicks  the ``OK'' button (Screen C) and logs itself out.
\item
The logout leads to the entry screen (Screen D).
From this point, Ape has access to only a small number of functionalities (e.g., logging in).
Ape cannot log in due to the difficulty of auto-generating the username/password of the test account.
In the remaining 58 minutes, Ape explores two new activities in total.
\end{enumerate}

This example represents Exploration Space Partition described in \S\ref{sec:intro}.
The essential problem is that Ape does not understand UI semantics---it does not know that the majority of OneNote's functionalities will be unreachable by clicking the ``OK'' button at the time of action.

\subsection{\TrappedLocal{}}

Figure~\ref{fig:ape_onenote_1}b  presents another example in which we run {Monkey}~\cite{AndroidMonkey},
a widely adopted tool, to test another popular industry-quality app, \textit{Nike Run Club}.
In this example, Monkey spends about 22 minutes trying to saturate one of the app's functionalities.
After investigating into the collected UI trace, we find the following behavior when Monkey interacts with the app:

\begin{enumerate}[leftmargin=*]
\item
Monkey explores other functionalities normally before entering Screen E that allows the tool to enter the functionality where the tool later gets trapped.
We name the functionality \textit{the trapping functionality}.
Monkey clicks the ``START'' button and enters the trapping functionality (Screen F).
\item
Monkey keeps clicking around in the trapping functionality.
To escape from the trapping functionality, Monkey first needs to press the Back button, and a confirmation dialog (Screen G) will pop up.
Monkey then has to click the ``OK'' button to finish escaping.
However, due to being widget-oblivion, Monkey  clicks only randomly on the screen, resulting in constant failures to click ``OK'' when the confirmation dialog is shown.
Furthermore, the dialog disappears when Monkey clicks outside of its boundary, and Monkey needs to press the Back button again to make the confirmation show up one more time.
It takes 22 minutes for Monkey to find and execute an effective escaping UI event sequence and finally leave the trapping functionality.
\item
The aforementioned behaviors are repeatedly observed in the trace (with different amounts of time used for escaping).
\end{enumerate}

This example represents \TrappedLocal{} behavior described in \S\ref{sec:intro}.
The essential problem is that Monkey is both widget- and state-oblivion, i.e., the tool is unable to locate actionable UI elements efficiently or sense whether it has been trapped and react accordingly (e.g., by restarting the target app).

\subsection{Implications}

To prevent such undesirable exploration behaviors, a conceptually simple idea is to de-prioritize exploring the entries to aforementioned trapping states (i.e., the ``OK'' button in Screen C, and ``START'' button in Screen F).
One potential solution is to develop natural language processing (NLP) or image processing based approaches that can infer the semantics of UI elements~\cite{10.1145/3319535.3363193,10.1109/ICSE.2019.00041,10.1145/3242587.3242650,Packevicius2018TextSA}.
While solutions based on understanding UI semantics are revolutionary, they are challenging due to fundamental difficulties rooting in NLP and image processing.

In this paper, we explore a more practical and evolutionary solution based on understanding \expNames{} by mining UI traces.
We show that it is feasible to identify the existence and location of such behavior through pattern analysis on interaction history.
Given the location of \expNames{}, we can further identify which UI actions might have led to such behavior.
Taking the example of Figure~\ref{fig:ape_onenote_1}, Ape starts to visit a very different set of screens (e.g., the welcome screens in Screen D in Figure \ref{fig:ape_onenote_1}a) after clicking ``OK'', and the number of explored screens dramatically decreases.
Therefore, we can look at the screen history and find the time point where the symptom starts to appear.
The UI action located at the aforementioned time point is then likely the cause of the symptom.
Our \ourtool{} system uses a specialized algorithm (\S\ref{sub:logoutpattern}) to effectively locate the starting time point of \expNames{} similar to the aforementioned instance.

\section{Background}

This section presents background knowledge about UI hierarchy to help readers understand our algorithm design and implementations in the scope of Android UI testing.

A \textit{UI hierarchy} structurally represents the contents of app UI shown at a time.
Each UI hierarchy consists of \textit{UI properties} (e.g., location, size) for individual \textit{UI elements} (e.g., buttons, textboxes) and hierarchical relations among UI elements.
On Android, each activity internally maintains the data structure for its current UI hierarchy.
Typically, UI elements are represented by \texttt{View}~\cite{AndroidView} subclass instances, and hierarchical relations are represented by child \texttt{View}s of \texttt{ViewGroup}~\cite{AndroidViewGroup} subclass instances.

\label{sub:ui_hierarchy_eqv_chk}

A key component of UI testing is to identify the current app functionality. 
The functionality is identified  by {\it equivalence check for UI hierarchies}, because UI hierarchies are usually used as indicators of apps' functionality scenarios.
Thus, checking the equivalence between the current and past UI hierarchies allows tools to identify whether a new functionality is being exercised.
If the current functionality has been covered, the tool can additionally leverage the knowledge associated with the functionality to decide on the next actions.
There are different ways to check UI hierarchy equivalence:

\begin{itemize}[leftmargin=*]
\item {\bf Strict comparison.}
A simple way to check the equivalence of two UI hierarchies is to compare their UI element trees and see whether they have identical structure and UI properties at each node.
In practice, such simple equivalence checking is too strict.
For example, on an app accepting text inputs, a tool checking exact equivalence can count a new functionality every time one character is typed.

\item {\bf Checking similarity.}
A workaround to the aforementioned issue of strict comparison is to check similarities of two UI hierarchies against a threshold.
However, ambiguity can become the new issue, given that the similarity relation is not transitive: suppose that A is similar to both B and C, it is still possible that B is not similar to C.
Then if both B and C are in the history (regarded as different functionalities), and A comes as a new UI hierarchy, the tool is  unable to decide on which functionality to use the associated knowledge from.
To fix the ambiguity issue, we can perform \textit{screen clustering}, essentially putting mutually similar screens into individual groups and regarding each group as representing one single functionality.
Then the downside is that screen clustering can be a computationally expensive operation, especially for traces with many screens.

\item {\bf Comparing abstractions.}
A more advanced solution is to check the equivalence at an abstraction level, employed by many model-based UI testing tools~\cite{Choi:2013,Gu:2019:PGT:3339505.3339542,su2017guided,10.1145/2970276.2970313,AimDroid}.
In the previous example, one can leave out all user-controlled textual UI properties from the hierarchy and the equivalence check can tell that the tool is staying on the same screen regardless of what has been entered.
While abstracting UI hierarchies is conceptually effective, it is challenging to design effective UI abstraction functions.
The difficulty lies in identifying UI properties or structural information to distinguish different app functionalities, especially when screens have variants with relatively subtle differences. %

$\quad$Ape~\cite{Gu:2019:PGT:3339505.3339542} includes adaptive abstractions to address the challenge of automatically finding proper UI abstraction functions in different scenarios.
Ape dynamically adjusts its abstraction strategy (e.g., which UI property values should be preserved) during testing based on feedback from strategy execution (e.g., whether invoking actions on UI hierarchies with the same abstraction yields the same results).
Unfortunately, the adaptive abstraction idea assumes the availability of sufficiently diversified execution history for feedback, and such history is not always available when analyzing given traces as in our situation.
\end{itemize}

Given the pros and cons of the aforementioned ways, we empirically adopt a hybrid approach for UI hierarchy equivalence check.
First, we always abstract UI hierarchies:
(1) we consider only visible UI elements (i.e., \texttt{View.getVisibility() == VISIBLE}, and the element's bounding box intersects with its parent's screen region),
(2) we keep the activity ID and the original hierarchical relations among UI elements, and
(3) we retain only UI element types and IDs from UI properties.
Second, we check the similarities of abstract UI hierarchies and cluster them into groups only when the analysis is sensitive to the absolute number of distinct screens.
More details on achieving clustering efficiently are elaborated in \S\ref{sub:trappedlocal}.

\section{The \ourtool{} Approach}

\label{sec:algorithms}

\subsection{Overview}
\label{sec:approach_overview}

\begin{figure}[t]
\centering
\includegraphics[width=0.85\columnwidth]{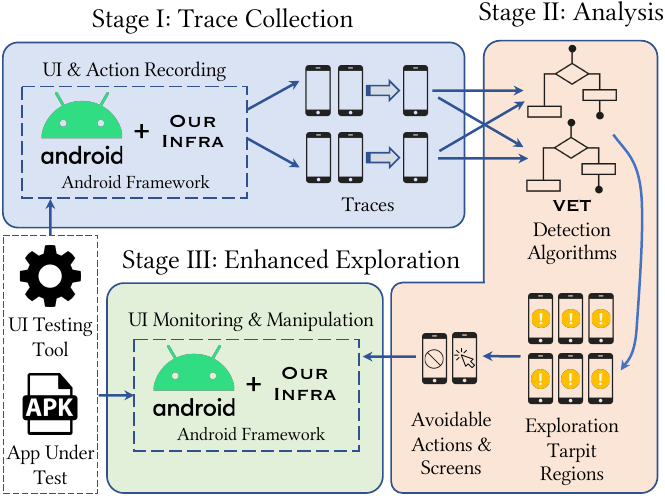}
\caption{Overview of \ourtool{}.}
\Description[An overview figure of \ourtool{}.]
{An overall illustration of how \ourtool{} works on the high level.}
\label{fig:vet_overview}
\end{figure}

We propose \ourtool{}, a general approach and its supporting system that automatically identifies and addresses \expNames{} for \textit{any} given Android UI testing tool on \textit{any} given AUT.
Our implementation of \ourtool{} is publicly available at~\cite{vet-github}.

As illustrated in Figure \ref{fig:vet_overview}, for a given tool and AUT, \ourtool{} works in three stages.
First, \ourtool{} runs the target tool on the AUT for a certain amount of time and records the interactions between the tool and AUT.
With help from our Android framework extension \toller{}~\cite{toller-paper}, \ourtool{} collects \textit{trace}(s) that consist of AUT UIs interleaving with the tool's actions.
Then, \ourtool{} analyzes each individual trace with specialized algorithms to identify trace subsequences (termed {\it regions}) that manifest the tool's \expNames{}.
Optionally, one can rank the identified regions based on their time lengths, where longer regions receive higher ranks, to prioritize regions that are likely to exhibit \expNames{} with higher impacts (see \S\ref{sub:root_causes_of_ineffective_exploration_intervals}).
Finally, \ourtool{} learns from the identified regions and guides the tool in subsequent runs to avoid \expNames{}, by monitoring the testing progress and taking actions based on findings from the identified regions.
With the support from \toller{}, \ourtool{} is currently capable of (1) preventing specified actions by disabling the corresponding UI elements at runtime and (2) assisting the AUT to escape from the specified screens by restarting the AUT.
The identified regions additionally support manual investigations of testing efficacy by providing localization help.

We equip \ourtool{} with two specialized algorithms targeting two  patterns of \expNames{}: \textit{\ExpSpacePartition{}} and \textit{\TrappedLocal{}}.
Characteristics of the two algorithms' targeted patterns are illustrated in Figure~\ref{fig:pattern_illust} and discussed as follows:

\begin{figure}[t]
\centering
\includegraphics[width=\columnwidth]{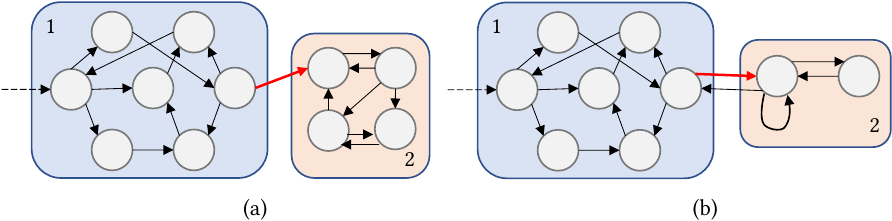}
\begin{justify}
\small{Note that each subgraph corresponds to an example trace, where each circle represents a distinct screen in the trace (e.g., each subfigure in Figure \ref{fig:ape_onenote_1}), each arrow indicates that action(s) is observed between two screens in the trace, and each curved rectangle depicts a UI subspace.
Red arrows denote destructive actions (e.g., clicking ``OK'' in Screen C of Figure \ref{fig:ape_onenote_1}a, ``START'' in Screen E of Figure \ref{fig:ape_onenote_1}b) while dashed arrows show where traces begin.}
\end{justify}
\caption{Two patterns of \expNames{}.}
\Description[Illustration of the two patterns of \expName{}.]
{Two transition graphs depicting \ExpSpacePartition{} and \TrappedLocal{}, respectively.}
\label{fig:pattern_illust}
\end{figure}

\begin{itemize}[leftmargin=*]
\item  \textit{\ExpSpacePartition{}}.
As shown in Figure \ref{fig:pattern_illust}a, the UI testing tool traverses through a UI subspace (Subspace 2) for a long time after the execution of some action (the red arrow), and the tool is \textit{unable} to return to the previously visited UI subspace (Subspace 1).
Furthermore, the tool visits much fewer distinct screens after the action.
The presence of the symptom suggests that the tool has triggered a destructive action  (effectively the partition boundary of the entire trace and beginning of the \expName{}) that prevents the tool from further exploring the app's major functionalities.
The first motivating example from \S\ref{sec:example} corresponds to this symptom, where clicking the ``OK'' button is the destructive action that gets Ape trapped in multiple screens related to logging in (Subspace 2) and prevents Ape from further accessing OneNote's main functionalities (Subspace 1).

\item \textit{\TrappedLocal{}}.
As shown in Figure \ref{fig:pattern_illust}b, the UI testing tool is trapped in a small UI subspace (Subspace 2) for an extended amount of time after the execution of the corresponding destructive action (the red arrow).
However, the tool is \textit{capable} of returning to the previously visited UI subspace (Subspace 1) despite the difficulties.
It is also likely that the tool will get trapped again within Subspace 2 after returning to Subspace 1.
Consequently, the tool spends an excessive amount of time repetitively testing limited functionalities in this hard-to-escape subspace.
The second motivating example from \S\ref{sec:example} corresponds to this symptom, where clicking the ``START'' button gets Monkey trapped in Screens F and G (Subspace 2).
Clicking ``OK'' helps Monkey go back to Screen E (within Subspace 1) and other functionalities, but it does not take a long time before the tool gets trapped again within Subspace 2.
\end{itemize}

As can be seen, \ExpSpacePartition{} targets  higher-level irreversible transition of UI exploration space, while \TrappedLocal{} focuses on lower-level difficulties of exercising a specific functionality.
Note that it is possible for the regions reported by the two algorithms on the same trace to overlap.
For example, \TrappedLocal{} might also capture \expNames{} within \ExpSpacePartition{}'s trapped UI subspace (corresponding to Subspace 2 in Figure \ref{fig:pattern_illust}).
Such overlaps do not prevent us from finding meaningful targeted \expNames{}: different \expNames{} revealed by regions identified by both algorithms suggest the existence of different exploration difficulties.

In the remaining of this section, we describe the two algorithms for capturing \ExpSpacePartition{} and \TrappedLocal{} in \S\ref{sub:logoutpattern} and \S\ref{sub:trappedlocal}, respectively.
We show that pattern capturing can be expressed as optimization problems.
Table~\ref{tab:notation} describes the notations used to describe \ourtool{}'s algorithms.

\begin{table}[t]
\small
\centering
\caption{Notations and descriptions used in the algorithms}
\begin{tabular}{ll}
\toprule
{\it Notation} & {\it Description} \\
\midrule
$S_i$        & Screen $\#i$ in the trace represented by the UI hierarchy. \\
$t_i$        & The timestamp of screen $S_i$ being observed.\\
$S_{l,r}$     & A region of screens starting at Screen $\#l$ and ending at \\
             & Screen $\#r$ (with both ends included). \\
$\{S_{l,r}\}$ & The set of distinct screens from $S_{l,r}$ by de-duplicating \\
             & their UI hierarchies.\\
$|S^s_{l,r}|$ & The number of occurrences of $s$ in $S_{l,r}$. \\
$t_{\min}$   & A predefined threshold that decides the minimum time \\
             & length of any $S_{l,r}$ (i.e., $t_r-t_l\geqslant t_{\min}$) that may be \\
             & included in algorithm outputs.\\
\bottomrule
\end{tabular}
\label{tab:notation}
\end{table}

\subsection{Capturing \ExpSpacePartition{}}
\label{sub:logoutpattern}

According to our introductions of \ExpSpacePartition{}, we need to find a destructive action exerted on screen $S_n$ as the partition boundary such that the aforementioned characteristics from \S\ref{sec:approach_overview} can be best reflected.
For instance, considering our first motivating example in \S\ref{sec:example}, we hope to pick up the screen shown in Figure \ref{fig:ape_onenote_1}c as $S_n$.
We optimize the following formula to find the most desirable $S_n$ from a trace with $N$ screens:

\begin{equation*}
\argmin_{1 \leqslant n < E_p}\quad
{[\!\sum\limits_{s\in\{S_{1,n}\}}\!\!\!\frac{|S_{n+1,N}^s|}{N-n}]} %
+ 2\cdot\sigma(\frac{|\{S_{n+1,N}\}|}{|\{S_{E_p+1,N}\}|}-1) - 1
\end{equation*}

In the formula, $E_p$ is a pre-calculated limit indicating the upper bound of $n$ during optimization, and $\sigma$ denotes the Sigmoid function. %
Note that $N-n$ can be pulled out of the sum subformula.
The intuition of the formula design is as follows:

\begin{enumerate}[leftmargin=*]
\item
As part of the characteristics, the tool should ideally be able to visit few to no screens that have appeared no later than $S_n$ after the tool passes $S_n$.
Correspondingly, in our motivating example, screens shown before Figure \ref{fig:ape_onenote_1}c (depicting the app's main functionalities) are dramatically different from the screens afterward (logging in, ToS, etc.).
In the formula, the nominator of the first term (intended to be minimized) quantifies the proportion of screens seen before $S_n$ within $S_{n+1,N}$. %

\item
As the denominator of the first term, $N-n$ essentially calculates how many (non-distinct) screens the tool visits after $S_n$.
There are two purposes of this design.
First, we hope to normalize the first term in the formula (so that two terms can weigh the same).
Given that $\sum_{s\in\{S_{1,n}\}}{|S_{n+1,N}^s|} = \sum_{s\in\{S_{1,n}\} \cap \{S_{n+1,N}\}}{|S_{n+1,N}^s|}$
$\leqslant\sum_{s\in\{S_{n+1,N}\}}{|S_{n+1,N}^s|} = N-n$,
 the first term is guaranteed to fall within $[0,1]$.
Second, we want to push $S_n$ backward (note that smaller $n$ makes the first term smaller) because we assume that the design makes $S_n$ closer to the \expName{}'s root cause, which should appear earlier than other causes.

\item
As another part of the characteristics, the tool stays within a certain UI subspace for a long time; thus, the tool will go through screens within the subspace very often.
If the tool generally uniformly visits most or all distinct screens within the subspace, by observing a small period of exploration (corresponding to $S_{E_p+1,N}$ in the formula) we should have a fairly precise estimation (i.e., $\{S_{E_p+1,N}\}$) of the subspace boundary, which is characterized by $\{S_{n+1,N}\}$.
The second term in the formula corresponds to this intuition, where the closer $\{S_{E_p+1,N}\}$ is to $\{S_{n+1,N}\}$ (note that $\{S_{E_p+1,N}\} \subseteq \{S_{n+1,N}\}$), the more favorable it becomes during optimization.

\item
By setting an upper bound $E_p$ on $n$ and regularizing the ratio with a Sigmoid function and applying appropriate linear transformations, we can guarantee that the second term in the formula always ranges from 0 to 1, being the same as the first term.
In the end, two terms in the formula contribute equally to optimization choices.
\end{enumerate}

To determine $E_p$ on each trace, because our optimization scope does not include any interval shorter than $[E_p,N]$, we choose a value such that $t_N - t_{E_p}$ is closest to $t_{\min}$.

After obtaining a potentially suitable $S_n$ through optimizing the aforementioned formula, we additionally check whether $|\{S_{1,n}\}| > |\{S_{n+1,N}\}|$ is satisfied, essentially enforcing the property that the exploration space should be smaller after the partition.
Finally, the reported region is $S_{n+1,N}$.

\subsection{Capturing \TrappedLocal{}}
\label{sub:trappedlocal}

Based on the characteristics of \TrappedLocal{} from \S \ref{sec:approach_overview}, we should track the presence of a region showing that the tool is trapped within a small UI subspace for an extended amount of time.
For our second motivating example in \S\ref{sec:example}, one valid choice is the 22-minute region starting from the button click in Screen E of Figure \ref{fig:ape_onenote_1}b.
We accordingly optimize the following formula to find the boundaries $S_l$ and $S_r$ of the most suitable region on a trace:

\begin{equation*}
\argmin_{1 \leqslant l \leqslant r \leqslant N}
\frac{|\textsc{Merge}(\{S_{l,r}\})|}{r-l+1}
\end{equation*}

In the formula, \textsc{Merge} denotes the operation of merging similar screens and returning the groups of merged screens.
As the optimization formula suggests, we hope to find a suitable region such that it covers few distinct screen groups despite that the tool tries to explore diligently (by injecting numerous actions quantified by $r-l+1$).
Then if $t_r - t_l \geqslant t_{\min}$, we regard that the \expName{} region $S_{l,r}$ can be reported.
Accordingly in our motivating example, the choice of $S_l$ is Screen F of Figure \ref{fig:ape_onenote_1}b and $S_r$ is the last instance of Screen G of Figure \ref{fig:ape_onenote_1}b in the 22-minute region.
$S_{l-1}$ corresponds to Screen E of Figure \ref{fig:ape_onenote_1}b, and the destructive action is reported.

Note that there can be more than one region exhibiting \TrappedLocal{} behavior within a single trace, given the possibility for the tool to escape the UI subspaces where \TrappedLocal{} behavior is observed.
In order to find all potential regions, we iterate the aforementioned optimization process on remaining region(s) each time after one region is chosen, until no more region can be divided.

\begin{algorithm}[t]
{\small
\KwData{A set of abstract UI hierarchies $H$}
\KwResult{A mapping $R: H \mapsto R$, where $R \subseteq H$}
Sort $h\in H$ by $|h|$ in ascending order \\
$R \leftarrow \{\}$ \\
\ForEach {$h \in H$} {
    $R[h] \leftarrow$ nil
}
\ForEach {$h \in H$} {
    \If {$R[h] =$ nil} {
        $R[h] \leftarrow h$ \\
        \ForEach {$h' \in H$} {
            \If {\textnormal{$R[h'] =$ nil $\land$ \textsc{SimCheck}$(h,h')$}} {
                $R[h'] \leftarrow h$
            }
        }
    }
}
\Return $R$
}
\caption{{\textsc{Merge}: Merging similar screens into groups (each group is represented by its root screen in $R$)}}
\label{algo:merge}
\end{algorithm}

\textbf{Design of \textsc{Merge}.}
As mentioned in \S \ref{sub:ui_hierarchy_eqv_chk}, involving screen merging is especially useful for handling \TrappedLocal{}, given that the aforementioned optimization formula is very sensitive to the absolute numbers of distinct screens.
Being part of the challenge, an efficient (and mostly effective) screen merging algorithm requires careful design.
Given a set of distinct abstract screens (represented by UI hierarchies) to merge, a relatively straightforward (and precise) approach is to first calculate the tree editing distance~\cite{tree_edit_dist} for each pair of abstract UI hierarchies for similarity check, and then use combinatorial optimization~\cite{wolsey1999integer} to decide the optimal grouping strategy (e.g., by converting to an Integer Linear Programming~\cite{linear_programming} problem), such that all screens within the same group are mutually similar and the total number of groups is minimal.
Unfortunately, such an algorithm requires exponential time in regards to the number of distinct abstract screens to merge.
The design will likely fall short on traces collected using industrial-quality apps, from which we can easily capture hundreds to thousands of distinct abstract screens.

Aiming to make the algorithm practically efficient, we relax the definition of similarity and the goal of optimization from the aforementioned merging algorithm based on insights from our observations.
Specifically, we find that in many cases, similar screens can be seen as screen variants derived from base screens by inserting a small number of leaf nodes or subtrees into the abstract UI hierarchy.
Based on this assumption with some tolerance for inaccuracy, we can
(1) design a more efficient tree similarity checker (Algorithm \ref{algo:ui_sim_check}), which considers only node insertion distances and has $\Theta(|h_1|\cdot|h_2|)$ time complexity (compared with $O(|h_1|\cdot|h_2|\cdot\textsc{Height}(h_1)\cdot\textsc{Height}(h_2))$ for full tree edit distance), and
(2) replace the inefficient combinatorial optimization with a highly efficient greedy algorithm (Algorithm \ref{algo:merge}), which tries to find all the base screens with $O(|H|^2\cdot\max_{h\in H}|h|^2)$ time complexity.
In practice, with multiple other optimizations not affecting the level of time complexity, the algorithm needs only several seconds on average to process a trace.
Even for a very long trace with 2,000 distinct abstract screens and tens of thousands of concrete screens, the algorithm runs for only several minutes.

\begin{algorithm}[t]
{\small
\KwData{Abstract UI hierarchies $h_1,h_2$}
\KwResult{Whether $h_1,h_2$ are similar enough}
\textbf{Const:} Max allowed distance $d_{\text{max}}$, empirically set to $3$ \\
$seq_1 \leftarrow []$ \\
\ForEach {$node\in$ \textnormal{\textsc{DepthFirstTraverse}}$(h_1)$} {
    $seq_1 \leftarrow seq_1 :: (\textsc{Props}(node), \textsc{Depth}(node))$ \\
    // \textsc{Props} obtains the node's UI properties that are preserved during abstraction. See more details in  \S\ref{sub:ui_hierarchy_eqv_chk}.
}
$seq_2 \leftarrow []$ \\
\ForEach {$node\in$ \textnormal{\textsc{DepthFirstTraverse}}$(h_2)$} {
    $seq_2 \leftarrow seq_2 :: (\textsc{Props}(node), \textsc{Depth}(node))$
}
$lcs \leftarrow \textsc{LongestCommonSequence}(seq_1, seq_2)$ \\
\eIf {$|lcs| < \min(|h_1|,|h_2|)$} {
    \Return \texttt{false}
} {
    \Return $\max(|h_1|,|h_2|) - |lcs| \leqslant d_{\text{max}}$
}
}
\caption{\textsc{SimCheck}: UI hierarchy similarity checker}
\label{algo:ui_sim_check}
\end{algorithm}

\section{Evaluation}

Our evaluation answers the following research questions:

\begin{itemize}[leftmargin=*]
\item \textbf{RQ1:}
How effectively can \ourtool{} help reveal Android UI testing tool issues with the identified \expName{} regions?
\item \textbf{RQ2:}
What is the extent of effectiveness improvement of Android UI testing tools through automatic enhancement by  \ourtool{}?
\item \textbf{RQ3:}
How likely do \ourtool{} algorithms miss tool issues in their identified \expName{} regions?
\end{itemize}

\subsection{Evaluation Setup} %
\label{sub:evaluation_setup}

{\bf Android UI Testing Tools and Android Apps.}
We use three state-of-the-art/practice Android UI testing tools:
Monkey~\cite{AndroidMonkey}, Ape~\cite{Gu:2019:PGT:3339505.3339542}, and WCTester~\cite{Zeng:2016,zheng17:automated}. %
We use \numofapps{} popular industry Android apps from the Google Play Store, as shown in Table \ref{tbl:ind_apps_details}. %
These \numofapps{} apps are from a previous study~\cite{Wang:2018:ESA:3238147.3240465}, which picks the most popular apps from each of the categories on Google Play and compares multiple testing tools applied on these apps.
The apps that we choose need to work properly on our testing infrastructure:
(1) they need to provide x86/x64 variants of native libraries (if they have any),
(2) they do not constantly crash on our emulators,
and (3) \toller{} is able to obtain UI hierarchies on most of the functionalities.
We additionally skip apps that
(1) have relatively limited sets of functionalities,
or (2) require logging in for access to most features and we are unable to obtain a consistently usable test account (e.g., some apps have disabled our test accounts after some experiments).

\begin{table}[t]
\begin{threeparttable}
\centering
\footnotesize
\caption{Overview of industrial apps used for evaluation}
\label{tbl:ind_apps_details}
\begin{tabular}{|l|r|c|r|c|}
\hline
{\textbf{App Name}} &
{\textbf{Version}} &
{\textbf{Category}} &
{\textbf{\#Inst}} &
{\textbf{Login}}
\\
\hline
AccuWeather	& 	5.3.5-free	& 	Weather	& 	50m+	& 	\xmark \\
AutoScout24	& 	9.3.14	& 	Auto \& Vehicles	& 	10m+	& 	\xmark \\
Duolingo	& 	3.75.1	& 	Education	& 	100m+	& 	\xmark \\
Flipboard	& 	4.1.1	& 	News \& Magazines	& 	500m+	& 	\cmark \\
Merriam-Webster	& 	4.1.2	& 	Books \& Reference	& 	10m+	& 	\xmark \\
Nike Run Club	& 	2.14.1	& 	Health \& Fitness	& 	10m+	& 	\cmark \\
OneNote	& 	16.0.9126	& 	Business	& 	100m+	& 	\cmark \\
Quizlet	& 	3.15.2	& 	Education	& 	10m+	& 	\cmark \\
Spotify	& 	8.4.48	& 	Music \& Audio	& 	100m+	& 	\cmark \\
TripAdvisor	& 	25.6.1	& 	Food \& Drink	& 	100m+	& 	\cmark \\
trivago	& 	4.9.4	& 	Travel \& Local	& 	10m+	& 	\xmark \\
Wattpad	& 	6.82.0	& 	Books \& Reference	& 	100m+	& 	\cmark \\
WEBTOON	& 	2.4.3	& 	Comics	& 	10m+	& 	\xmark \\
Wish	& 	4.16.5	& 	Shopping	& 	100m+	& 	\cmark \\
YouTube	& 	15.35.42	& 	Video Player \& Editor	& 	1b+	& 	\xmark \\
Zedge	& 	7.2.2	& 	Personalization	& 	100m+	& 	\xmark \\
\hline
\end{tabular}
\begin{tablenotes}
\small
\item \textit{Notes:} `\#Inst' denotes the approximate number of downloads. `Login' indicates whether the app requires logging in to access most features.
\end{tablenotes}
\end{threeparttable}

\end{table}

{\bf Trace Collection.}
We run each tool on every app for three times to alleviate the potential impacts of non-determinism in testing.
Each run takes one hour without interruption, and we restart the tool if it exits before using up the allocated run time.
\toller{} records one UI trace for each test run.
While \ourtool{} runs separately on each UI trace, results are grouped for each (tool, app) pair.
In total, we collect \numoftraces{} one-hour UI traces from \numofcombs{} (tool, app) pairs.

{\bf Testing Platform.}
All experiments are conducted on the official Android x64 emulators running Android 6.0 on a server with Xeon E5-2650 v4 processors.
Each emulator is allocated with 4 dedicated CPU cores, 2 GiB of RAM, and 2 GiB of internal storage space.
Emulators are stored on a RAM disk and backed by discrete graphics cards for minimal mutual influences caused by disk I/O bottlenecks and CPU-intensive graphical rendering.
We manually write auto-login scripts for apps with ``Login'' ticked in Table \ref{tbl:ind_apps_details}, and each of these scripts is executed only once before the corresponding app starts to be tested in each run.
To alleviate the flakiness of these auto-login scripts, we manually check the collected traces afterward and rerun the experiments with failed login attempts.

{\bf Overall Statistics.}
\ourtool{} reports \numOfRegions{} regions to exhibit \expNames{}, averaging 2.7 on each (tool, app) pair.
Based on \ourtool{}'s reports, the average amount of time involved in \expNames{} is \avgRegionLengthMinutes{} minutes per region, with the maximum being 59 minutes and minimum being slightly more than 10 minutes (given that we empirically set $t_{\min} = 10$ minutes for all the  experiments).

\subsection{RQ1. Detected Tool Issues} %
\label{sub:root_causes_of_ineffective_exploration_intervals}
\subsubsection{Methodology}
\label{sub:manual_eval_methodology}

We evaluate the effectiveness of \ourtool{} algorithms in capturing \expNames{} that reveal issues of testing tools upon AUTs.
Specifically, we first group \expName{} regions by the (tool, app) pairs that these regions are observed on.
Then, we rank the regions within each (tool, app) pair by their time lengths as mentioned in \S\ref{sec:approach_overview}.
Finally, we manually investigate each of \numOfRegions{} regions from all (tool, app) pairs.
We report any issue for each of these regions with manual judgment.
Note that we count only the issue that we consider most specific to the \expName{} revealed by each region:
  if both issues A and B contribute to the \expName{} on some region, and A also contributes to other regions on the same trace, we count only B in the statistics.

\subsubsection{Results}
We are able to determine tool issues on \numofcombswithissue{} of \numOfRegions{} manually investigated regions.
Table \ref{tbl:issue_type_dist} shows the distribution of issue types w.r.t the tool and region ranking.
Note that we find each (tool, app) pair to have up to three regions reported by \ourtool{};
  thus, rank-1/2/3 regions cover all \numOfRegions{} regions (with 48/43/40 regions each).
The tool issues can be traced to two root causes:
\textit{apps under test require extra knowledge for effective testing}, and
\textit{tool defects prevent themselves from progressing.}
We discuss specific issues w.r.t. these two root causes: 

\begin{table}[t]
\begin{threeparttable}
\centering
\setlength{\tabcolsep}{2pt}
\footnotesize
\caption{Distribution of confirmed issue types}
\label{tbl:issue_type_dist}
\begin{tabular}{|c|rrrr|rrrr|rrrr|r|} \hline
\multirow{2}{*}{\textbf{Issue}}
& \multicolumn{4}{c|}{\textbf{Rank-1}} & \multicolumn{4}{c|}{\textbf{Rank-2}} & \multicolumn{4}{c|}{\textbf{Rank-3}} &
\multirow{2}{*}{\textbf{Total}}
\\ \cline{2-13}
& \multicolumn{1}{c}{Ape} & \multicolumn{1}{c}{Mk} & \multicolumn{1}{c}{Wt} & \multicolumn{1}{c|}{Sum} & \multicolumn{1}{c}{Ape} & \multicolumn{1}{c}{Mk} & \multicolumn{1}{c}{Wt} & \multicolumn{1}{c|}{Sum} & \multicolumn{1}{c}{Ape} & \multicolumn{1}{c}{Mk} & \multicolumn{1}{c}{Wt} & \multicolumn{1}{c|}{Sum} &
\\ \hline
\textbf{LOUT} & 5     & 2     & 1     & 8     & 5     & 2     & 1     & 8     & 4     & 1     & 1     & 6     & \textbf{22} \\
\textbf{UI} & 0     & 4     & 0     & 4     & 0     & 3     & 0     & 3     & 2     & 4     & 0     & 6     & \textbf{13} \\
\textbf{NTWK} & 0     & 5     & 0     & 5     & 0     & 3     & 0     & 3     & 0     & 3     & 0     & 3     & \textbf{11} \\
\textbf{LOOP} & 0     & 0     & 7     & 7     & 0     & 0     & 8     & 8     & 0     & 0     & 5     & 5     & \textbf{20} \\
\textbf{ESC} & 0     & 4     & 0     & 4     & 0     & 2     & 0     & 2     & 0     & 2     & 0     & 2     & \textbf{8} \\
\textbf{ABS} & 0     & 0     & 2     & 2     & 0     & 0     & 2     & 2     & 0     & 0     & 1     & 1     & \textbf{5} \\
\textbf{MISC} & 5     & 1     & 2     & 8     & 5     & 1     & 0     & 6     & 3     & 0     & 0     & 3     & \textbf{17} \\ \hline
\textbf{Total} & \textbf{10} & \textbf{16} & \textbf{12} & \textbf{38} & \textbf{10} & \textbf{11} & \textbf{11} & \textbf{32} & \textbf{9} & \textbf{10} & \textbf{7} & \textbf{26} & \textbf{96} \\ \hline
\end{tabular}
\begin{tablenotes}
\small
\item \textit{Notes:}
`Mk' and `Wt' refer to Monkey and WCTester, respectively.
Issue type details are discussed in \S\ref{sub:root_causes_of_ineffective_exploration_intervals}.
\end{tablenotes}
\end{threeparttable}

\end{table}

\begin{itemize}[leftmargin=*]
\item {\bf App logout or equivalent} (abbreviated as ``LOUT'' in Table \ref{tbl:issue_type_dist})
accounts for \expNames{} on 23\% (22 out of \numofcombswithissue{}) of investigated regions with identified issues.
Some apps essentially require login states for the majority of their functionalities to be accessible.
However, the tools used in our experiments have no knowledge about the consequences of clicking the ``logout'' buttons in different apps before the tools actually try clicking these buttons.
Unfortunately, after the tools try out such actions (driven by their exploration strategies), the apps' login states (both on-device and on-server) have been destroyed.
The tools then have to spend all remaining time on a limited number of functionalities, leading to \expNames{}.
This case reveals a common weakness of existing UI testing tools---they have a limited understanding of action semantics.
As can be seen from Table \ref{tbl:issue_type_dist}, Ape is more likely to be affected by this type of issues.

\item {\bf Unresponsive UIs} (``UI'') are found in 14\% (13 out of \numofcombswithissue{}) of regions.
We find that some apps stop responding to UI actions after the advertisement banner is clicked, even though the apps' UI threads are not blocked.
The issue is likely caused by the UI design defects in the Google AdMob SDK.
The AUTs should be restarted as soon as possible to resume access to their functionalities.
According to Table \ref{tbl:issue_type_dist}, Monkey is most vulnerable to such issues.
One interesting finding is that Ape is actually also vulnerable to unresponsive UIs although the tool's implementation is capable of identifying such situations.
However, while Ape proceeds to restart the app most of the times when Ape finds the app unresponsive, Ape fails to do so occasionally.

\item {\bf Network disconnections} (``NTWK'') are found in 11\% (11 out of \numofcombswithissue{}) of regions.
We find that turning networking off is undesirable for some apps, especially when the disconnection lasts for a long time.
Consequently, these apps may show only messages prompting users to check their networks, leaving nothing for exploration.
Tools such as Monkey can control network connections through Android system UI (e.g., by clicking the ``Airplane Mode'' icon).
While the capability helps test app logic in edge conditions in general, it might hurt the tool's effectiveness on apps heavily relying on network access.
All regions with the aforementioned issue come from Monkey's traces.

\item {\bf Restart/action loops} (``LOOP'') are found in 21\% (20 out of \numofcombswithissue{}) of regions.
The tool essentially keeps restarting or performing the same actions on the target app after some point.
One potential cause for such issues is that the tool thinks that {\it all} UI elements in the target app's main screen have been explored.
The tool might need to revise its exploration strategy for discovering more explorable functionalities.

\item {\bf Obscure escapes} (``ESC'') are found in 8\% (8 out of \numofcombswithissue{}) of regions.
Defects in a tool's design or implementation can make it difficult for the tool to escape from certain app functionalities, and consequently, the tool loses opportunities to explore other functionalities.
For instance, Monkey finds it challenging to escape a screen where the only exit is a tiny button on the screen (see the second motivating example in \S\ref{sec:example}), due to the tool's lack of understanding of UI hierarchies.

\item {\bf UI abstraction defects} (``ABS'') are found in {5\%} (5 out of \numofcombswithissue{}) of regions.
Defects in a tool's UI abstraction strategies can trick the tool into incorrectly understanding the testing progress.
Seen from collected traces, WCTester considers all texts as part of abstract UI hierarchies.
While the strategy works well in a wide range of testing scenarios, it keeps the tool repetitively triggering actions on UI elements with changing texts (such as counting down), given that the tool incorrectly thinks that new functionalities are being covered.

\item {\bf Miscellaneous tool implementation defects} (``MISC'') are in 18\% (17 out of \numofcombswithissue{}) of regions.
In our case, we find potential implementation defects in three tools:
(1) Ape and Monkey fail to handle unresponsive apps (``Injection failed'' or being unable to obtain UI hierarchies),
and (2) WCTester is found to explore only a certain fraction of app functionalities after some point.
\end{itemize}

Another finding is that the ratio of confirmed issues decreases when the rank goes lower ($38/48=79\%$ for rank-1, $32/43=74\%$ for rank-2, and $26/40=65\%$ for rank-3), suggesting the usefulness of prioritizing regions based on their lengths.

\subsection{RQ2. Improvement of Testing Performance} %
\label{sub:integrating_our_algorithms_with_automated_ui_testing}
This section shows that the identified \expName{} regions by \ourtool{} can be used to automatically address tool issues.

\subsubsection{Automatic fix application}
The essential idea is to prevent some tool issues from happening again or getting rid of tool issues quickly by controlling the interactions between tools and apps.
Specifically, given an \expName{} region, we identify the UI element that the tool acts on right before the region begins, and then we use \toller{} to disable the UI element for \ourtool{}-guided runs.
Many tool issues can be targeted by this simple approach.
For example, if we disable the advertisement banners that lead to {Unresponsive UIs} in \S\ref{sub:root_causes_of_ineffective_exploration_intervals}, tools will simply not run into the undesirable situation, and they can focus on testing other more valuable app functionalities.
In some cases when there are multiple entries to the region and existing traces do not reveal all the entries, the aforementioned approach might fail.
We mitigate this limitation by monitoring and controlling the testing progress---currently, if we observe any of the most frequently appearing screens from \TrappedLocal{} regions, we restart the AUT in \ourtool{}-guided runs.

We implement UI element disablement by relying on \toller{} to monitor screen changes during testing and dynamically modify UI element properties.
When a target screen (i.e., a UI screen containing any target UI element, as determined by UI hierarchy equivalence check) shows up, we pinpoint the target UI element by matching with the path to each UI element from the root UI element.
Once we confirm that the target UI element exists on the current screen, we instruct \toller{} to disable the UI element, which will not respond to any further action on itself.
For the edge case where the action is not on any UI element (e.g., pressing the Back button), we restart the target app once we see the corresponding target screen.
Note that there is no need to modify the app installation packages given that we manipulate app UIs dynamically.

\subsubsection{Methodology of experiments}
We aim to measure the testing effectiveness throughout the entire process of applying \ourtool{}.
For each (tool, app) pair, in addition to the three {\it initial} runs for trace collection, we perform the experiments for three runs in each of these three settings:
(1) Disabling UI elements based on rank-1 regions
 (rank-1 \textit{guided} runs; see \S\ref{sec:approach_overview} for our ranking strategy),
(2) Disabling UI element based on rank-1, rank-2, and rank-3 regions (rank-1/2/3 guided runs), and
(3) Keep the same settings as initial runs (\textit{comparison} runs).
Note that each run also lasts for one hour, and all experiments are conducted in the same hardware and software environment regardless of different settings.

We measure method coverage (numbers of uniquely covered methods in app bytecode) as one testing effectiveness metric in our experiments.
Note that methods involved by app initialization (i.e., before tools start to test) are excluded for a more precise comparison of code coverage gain.
Upon each (tool, app) pair, we use the following Test Groups (TGs) for effectiveness comparison:

\begin{itemize}[leftmargin=*]
\item \textit{TG-1}: three initial runs and three comparison runs.
\item \textit{TG-2}: three initial runs and three rank-1 guided runs.
\item \textit{TG-3}: three initial runs and three rank-1/2/3 guided runs.
\end{itemize}

Each group consists of six one-hour runs, intended for reducing random biases.
We accumulate the method coverage of all runs within a group for the group's method coverage.
Test Group 1 serves as the baseline, while the other two test groups aim to measure how much testing effectiveness gain can \ourtool{} users expect.
The main reason for experimenting with \expName{} regions of different ranks is that there can be multiple tool issues on an AUT, and addressing only one of the issues might not suffice.

We also measure the crash triggering capabilities with cumulative numbers of distinct crashes.
We consider only crashes from bytecode given that Android apps are predominantly written in JVM languages (Java and Kotlin).
Crashes are identified by hashing the code locations in stack traces.
We additionally leverage \toller{} to disable each app's \texttt{UncaughtExceptionHandler}, which is widely used by industrial apps to collect crash reports and might prevent crash information from being exposed to the Android log system (i.e., Logcat~\cite{logcat}) and captured by our scripts.

It should also be noted that \toller{} monitors, captures, and manipulates UIs with negligible overheads; thus, the testing effectiveness of original runs should remain comparable with and without \toller{} in use.
In addition, \ourtool{} analyzes traces very efficiently, usually requiring only a few seconds on a single trace, while analysis of multiple traces can be trivially parallelized.

\begin{table*}[t]
\centering
\begin{threeparttable}
\footnotesize
\caption{Cumulative code coverage statistics.}
\label{tbl:intg_test_results}
\begin{tabular}{|l|r|rr|rr|r|rr|rr|r|rr|rr|}
\hline
\multirow{3}{*}{\textbf{App Name}} & 
\multicolumn{5}{c|}{\textbf{Ape}} &
\multicolumn{5}{c|}{\textbf{Monkey}} &
\multicolumn{5}{c|}{\textbf{WCTester}}
\\ \cline{2-16}
& \multirow{2}{*}{\textbf{\#M$_1$}}
& \multicolumn{2}{c|}{\textbf{Rank-1}} & \multicolumn{2}{c|}{\textbf{Rank-1/2/3}}
& \multirow{2}{*}{\textbf{\#M$_1$}}
& \multicolumn{2}{c|}{\textbf{Rank-1}} & \multicolumn{2}{c|}{\textbf{Rank-1/2/3}}
& \multirow{2}{*}{\textbf{\#M$_1$}}
& \multicolumn{2}{c|}{\textbf{Rank-1}} & \multicolumn{2}{c|}{\textbf{Rank-1/2/3}}
\\ \cline{3-6} \cline{8-11} \cline{13-16}
& & {\textbf{\#M$_2$}} & {\textbf{$\Delta$M$_2$}} & {\textbf{\#M$_3$}} & {\textbf{$\Delta$M$_3$}}
& & {\textbf{\#M$_2$}} & {\textbf{$\Delta$M$_2$}} & {\textbf{\#M$_3$}} & {\textbf{$\Delta$M$_3$}}
& & {\textbf{\#M$_2$}} & {\textbf{$\Delta$M$_2$}} & {\textbf{\#M$_3$}} & {\textbf{$\Delta$M$_3$}}
\\
\hline
AccuWeather & 21977 & 22485 & 2.3\% & 22538 & 2.6\% & 14830 & 29266 & 97.3\% & 24990 & 68.5\% & 14982 & 15104 & 0.8\% & 14797 & -1.2\% \\
AutoScout24 & 17245 & 17274 & 0.2\% & 22713 & 31.7\% & 23455 & 23270 & -0.8\% & 23085 & -1.6\% & 18637 & 22157 & 18.9\% & 22154 & 18.9\% \\
Duolingo & 15186 & 15299 & 0.7\% & 15510 & 2.1\% & 13676 & 14598 & 6.7\% & 14582 & 6.6\% & 12319 & 14625 & 18.7\% & 14450 & 17.3\% \\
Flipboard & 9594  & 9742  & 1.5\% & 9510  & -0.9\% & 5949  & 7482  & 25.8\% & 7125  & 19.8\% & 7872  & 7901  & 0.4\% & 8265  & 5.0\% \\
Merriam-Webster & 9056  & 9093  & 0.4\% & 9093  & 0.4\% & 5617  & 8992  & 60.1\% & 9275  & 65.1\% & 9328  & 9089  & -2.6\% & 9010  & -3.4\% \\
Nike Run Club & 30523 & 29937 & -1.9\% & 29939 & -1.9\% & 24472 & 24592 & 0.5\% & 26645 & 8.9\% & 19754 & 21936 & 11.0\% & 22460 & 13.7\% \\
OneNote & 6681  & 7134  & 6.8\% & 7028  & 5.2\% & 7131  & 7114  & -0.2\% & 7381  & 3.5\% & 6453  & 6675  & 3.4\% & 6933  & 7.4\% \\
Quizlet & 17000 & 17114 & 0.7\% & 16900 & -0.6\% & 13722 & 13995 & 2.0\% & 13995 & 2.0\% & 14679 & 14865 & 1.3\% & 14448 & -1.6\% \\
Spotify & 19759 & 21475 & 8.7\% & 21475 & 8.7\% & 20616 & 22200 & 7.7\% & 21486 & 4.2\% & 18897 & 29298 & 55.0\% & 19632 & 3.9\% \\
TripAdvisor & 29857 & 30645 & 2.6\% & 31006 & 3.8\% & 16773 & 20919 & 24.7\% & 20919 & 24.7\% & 26773 & 28467 & 6.3\% & 28180 & 5.3\% \\
trivago & 20706 & 20710 & 0.0\% & 20711 & 0.0\% & 20216 & 20489 & 1.4\% & 20482 & 1.3\% & 19952 & 19964 & 0.1\% & 20032 & 0.4\% \\
Wattpad & 22960 & 22668 & -1.3\% & 22447 & -2.2\% & 14541 & 13717 & -5.7\% & 15276 & 5.1\% & 15067 & 15884 & 5.4\% & 15982 & 6.1\% \\
WEBTOON & 32933 & 31674 & -3.8\% & 31599 & -4.1\% & 31477 & 30176 & -4.1\% & 30176 & -4.1\% & 25720 & 27659 & 7.5\% & 27659 & 7.5\% \\
Wish  & 8829  & 8850  & 0.2\% & 9106  & 3.1\% & 8490  & 8522  & 0.4\% & 8269  & -2.6\% & 6948  & 7191  & 3.5\% & 7207  & 3.7\% \\
Youtube & 26874 & 33301 & 23.9\% & 33757 & 25.6\% & 29316 & 32087 & 9.5\% & 35892 & 22.4\% & 22179 & 29143 & 31.4\% & 30233 & 36.3\% \\
Zedge & 42899 & 43074 & 0.4\% & 43433 & 1.2\% & 31245 & 38103 & 21.9\% & 44931 & 43.8\% & 36671 & 37464 & 2.2\% & 38343 & 4.6\% \\
\hline
\textbf{Average} & \textbf{20755} & \textbf{21280} & \textbf{2.5\%} & \textbf{21673} & \textbf{\covImprApe{}} & \textbf{17595} & \textbf{19720} & \textbf{12.1\%} & \textbf{20282} & \textbf{\covImprMonkey{}} & \textbf{17264} & \textbf{19214} & \textbf{\covImprWC{}} & \textbf{18737} & \textbf{8.5\%} \\
\hline
\end{tabular}
\begin{tablenotes}
\small
\item \textit{Notes:}
`\#M$_n$' shows the total number of covered methods in Test Group $n$.
$\Delta\text{M}_n = (\#\text{M}_n - \#\text{M}_1) \div \#\text{M}_1 \times 100\%$.
\end{tablenotes}
\end{threeparttable}

\end{table*}

\begin{table}[t]
\centering
\begin{threeparttable}
\setlength{\tabcolsep}{3pt}
\footnotesize
\caption{Distinct crash statistics.}
\label{tbl:intg_test_results_crashes}
\begin{tabular}{|l|rrr|rrr|rrr|}
\hline
\multirow{2}{*}{\textbf{App Name}} & 
\multicolumn{3}{c|}{\textbf{Ape}} &
\multicolumn{3}{c|}{\textbf{Monkey}} &
\multicolumn{3}{c|}{\textbf{WCTester}} \\ \cline{2-10}
& \multicolumn{1}{c}{\textbf{\#C$_1$}}
& \multicolumn{1}{c}{\textbf{\#C$_2$}} & \multicolumn{1}{c|}{\textbf{\#C$_3$}}
& \multicolumn{1}{c}{\textbf{\#C$_1$}}
& \multicolumn{1}{c}{\textbf{\#C$_2$}} & \multicolumn{1}{c|}{\textbf{\#C$_3$}}
& \multicolumn{1}{c}{\textbf{\#C$_1$}}
& \multicolumn{1}{c}{\textbf{\#C$_2$}} & \multicolumn{1}{c|}{\textbf{\#C$_3$}}
\\ \hline
AccuWeather & 2     & 4     & 9     & 0     & 5     & 4     & 0     & 2     & 4 \\
AutoScout24 & 1     & 1     & 1     & 2     & 1     & 1     & 0     & 0     & 0 \\
Duolingo & 1     & 2     & 2     & 0     & 0     & 0     & 1     & 1     & 1 \\
Flipboard & 0     & 0     & 1     & 0     & 0     & 1     & 1     & 1     & 0 \\
Merriam-Webster & 2     & 3     & 3     & 0     & 5     & 7     & 0     & 0     & 0 \\
Nike Run Club & 1     & 1     & 1     & 6     & 3     & 5     & 0     & 0     & 1 \\
OneNote & 0     & 2     & 5     & 0     & 2     & 1     & 0     & 1     & 0 \\
Quizlet & 0     & 1     & 0     & 0     & 0     & 0     & 1     & 1     & 1 \\
Spotify & 1     & 1     & 1     & 0     & 0     & 0     & 0     & 0     & 0 \\
TripAdvisor & 3     & 5     & 5     & 0     & 1     & 1     & 1     & 1     & 1 \\
trivago & 1     & 1     & 2     & 0     & 1     & 2     & 2     & 1     & 2 \\
Wattpad & 2     & 2     & 2     & 0     & 0     & 0     & 2     & 3     & 2 \\
WEBTOON & 1     & 1     & 1     & 1     & 0     & 0     & 0     & 3     & 3 \\
Wish  & 2     & 4     & 2     & 2     & 1     & 1     & 0     & 0     & 0 \\
Youtube & 0     & 0     & 0     & 0     & 0     & 0     & 1     & 1     & 2 \\
Zedge & 0     & 0     & 0     & 0     & 0     & 0     & 0     & 0     & 0 \\
\hline
\textbf{Total} & \textbf{17} & \textbf{28} & \textbf{35} & \textbf{11} & \textbf{19} & \textbf{23} & \textbf{9} & \textbf{15} & \textbf{17} \\ \hline
\end{tabular}
\begin{tablenotes}
\small
\item \textit{Notes:}
`\#C$_n$' is for total \# triggered unique crashes in Test Group $n$.
\end{tablenotes}
\end{threeparttable}

\end{table}

\subsubsection{Results}
Tables \ref{tbl:intg_test_results} and~\ref{tbl:intg_test_results_crashes} show the effectiveness improvements by comparing three test groups.
As can be seen from the results, automatically applying fixes based on \ourtool{}'s identified \expName{} regions helps Ape, Monkey, and WCTester achieve up to \covImprApe{}, \covImprMonkey{}, and \covImprWC{} cumulative code coverage improvements relatively on \numofapps{} apps using the same amount of time.
Additionally, \ourtool{} helps Ape, Monkey, and WCTester achieve up to \crashImprApe{}, \crashImprMonkey{}, and \crashImprWC{} overall distinct crashes, respectively.
It should be noted that \ourtool{}'s automatic approach does not address all the tool issues---some issues, especially those rooting in tool implementations, are likely  addressable by only humans.

For most (tool, app) pairs with improvements, considering only rank-1 \expName{} regions is sufficient for code coverage gain.
However, there are cases where code coverage increases considerably when we consider rank top-3 regions instead.
One explanation is that there are multiple applicability issues, or multiple instances of \expName{} corresponding to the same applicability issue.
For example, when applying Monkey on the app {Nike Run Club}, there are multiple ways to enter a hard-to-escape functionality as depicted by the motivating example in \S\ref{sec:example}.
If we block only one entry, Monkey can still find other ways to enter the functionality (despite being more difficult) and waste time there.

There are also cases where code coverage decreases when we consider rank-3 \expName{} regions.
One reason is that lower-ranked regions might not capture real issues, but \ourtool{} tries to ``fix'' them anyway, indeliberately interfering with normal functionalities.
In the case of applying WCTester on {Spotify}, the rank-3 region does not reveal any tool issue, according to our observation.
``Fixing'' this region can cause \ourtool{} to restart the app when one major functionality shows up.
Consequently, WCTester is unable to explore that functionality to achieve more coverage in guided runs.

\subsection{RQ3. Missed Tool Issues} %
We show our analysis of tool issues that are not revealed by any \expName{} regions (i.e., false negatives) reported by \ourtool{}.

\subsubsection{Methodology}

We propose using issue-specific detection tools to discover hidden tool issues (in all the collected traces), which can provide an estimation of how likely any issue is missed.
Specifically, we summarize the characteristics of two issues from \S\ref{sub:root_causes_of_ineffective_exploration_intervals}, \textit{App logout} and \textit{Unresponsive UIs}, to design two approaches specifically targeting these two issues.
The reason for choosing the aforementioned issues is that
(1) they have a substantial appearance among all issues that we have identified, and
(2) their existence is relatively more straightforward to be determined using our infrastructure.
We do not adopt manual inspection due to the subjectivity and the error-prone nature of manual judgments, especially given that we need to look at all the collected traces entirely.

Our issue-specific detection approaches work as follows:

\begin{itemize}[leftmargin=*]
\item \textit{App logout.}
We first manually look into the activity list of each app with `Login' ticked in Table \ref{tbl:ind_apps_details} and identify the subset of activities that are used for logging in to the app.
Then, when we analyze a given trace, we find the first and the last occurrence of any activity that belongs to the aforementioned list.
If there is any occurrence, and the time distance between the first and the last occurrence is at least $t_\text{min}$, we regard that an issue of {App logout} is found in the trace.

\item \textit{Unresponsive UIs.}
In our investigation, we find only one case that leads to {Unresponsive UIs}: when an advertisement banner is clicked.
Consequently, a new activity can be observed, where the activity belongs to the Google AdMob SDK and has the same activity ID across different apps.
Thus, we simply look for continuous appearance (i.e., there is no other activity in between) of the aforementioned activity ID on the given trace.
Note that we require the appearance to be continuous so as to exclude the cases where the tool (such as Ape) chooses to restart the app.
If the appearance lasts for at least $t_\text{min}$, we regard that an issue of {Unresponsive UIs} is found in the trace.
\end{itemize}

In order for a detected issue to be considered covered by our general-purpose algorithms, we require that at least one algorithm-identified \expName{} region covers at least 1/2 of the time length within which the detected issue appears.

\subsubsection{Results}

We apply the specialized approaches on all \numoftraces{} collected traces.
As a sanity check, we find that all the manually-discovered {App logout} and {Unresponsive UIs} issues are covered by the specialized approaches.
We compare the results against identified regions from \ourtool{}'s general-purpose algorithms and perform manual confirmation.
We find only several cases where the \ourtool{} algorithms do not yield accurate results, discussed as follows:

\begin{itemize}[leftmargin=*]
\item
On one trace from applying Monkey on Zedge, \ourtool{} misses one Unresponsive UIs issue by not reporting any covered region.
We find that the \TrappedLocal{} algorithm prioritizes another region over any region covering this issue.
However, we find that this issue is also present in other traces from the same (tool, app) pair, and \ourtool{} identifies and addresses this issue.

\item
On each of two other traces from Ape on Duolingo and Monkey on Zedge, there are two instances of the same Unresponsive UIs issue, and \ourtool{} reports only one of them.
The inaccuracy is also caused by the prioritization strategy of the \TrappedLocal{} algorithm.
However, since two instances point to the same issue on both traces, \ourtool{} is still effective.

\item
On one trace from Ape on Quizlet, Ape logs out about only 5 minutes after testing starts, but \ourtool{} reports only 22 minutes of \expName{}.
We find Quizlet's UI design to be somewhat unique: the app has a special entry to some of the main functionalities in its landing page that is accessible without logging in.
The entry is buried within a paragraph of texts, and the texts are shown only after a specific combination of swiping.
Ape is able to find this special entry in this run, making \ourtool{} confused.
Nevertheless, \ourtool{} is still able to find the correct trigger action from other traces.

\end{itemize}

\label{sub:characteristics_of_ineffective_testing}

\section{Discussion and Limitation}
We are mainly focusing on making \ourtool{} useful in the context of automated UI testing.
However, it should be noted that \ourtool{}'s potential usage scenarios are beyond automated UI testing.
One usage case is for app UI quality assurance, where an app might have UI design issues with one or more functionalities.
As a result, human users may face difficulties locating their desired features.
When a human user runs into such situations, he/she will then likely search for the desired features through repeated (and \ineffective{}) exploration around a few functionalities, and the difficulties can be reflected by the collected user behavior statistics.
By subsequently utilizing \ourtool{}'s identification of \expNames{}, we can quickly know which functionalities likely have the aforementioned UI design issues, potentially from numerous traces collected from end-users, and address these issues in a more timely manner.

One question is whether \ourtool{} is capable of differentiating testing scenarios
(1) that a tool is supposed to handle but does not (i.e., tool issues), and
(2) that a tool is not expected to handle (i.e., beyond tool capabilities, such as apps requiring special inputs).
We would like to point out that it is inherently difficult to differentiate these two types of scenarios due to lacking specifications over tool capabilities.
On the other hand, \ourtool{} can help users identify (and mitigate) the cases beyond a tool’s capabilities, being already useful.

We acknowledge that \toller{}, the utility that monitors, captures, and manipulates UIs for \ourtool{}, still has limitations.
For example, the current implementation of \toller{} does not capture text inputs.
However, adding support for text capturing is achievable with engineering efforts.
Moreover, \toller{}'s limitations do not prevent \ourtool{} from being generalizable.

\section{Threats of validity}
A major external threat to the validity of our work is the environmental dependencies of our subject apps.
More specifically, many of the industrial apps in our experiments require networking for main functionalities to be usable, and it is possible for such dependency to change the behaviors of these apps despite our efforts to make our experiment environment consistent across different runs.
In order to reduce the influences of environmental dependencies in our experiments, we repeat each experiment setting by three times and use aggregated metrics in our paper.
We additionally control each tool's internal randomness by setting a constant random seed for each app.
Nevertheless, this threat can be further reduced by involving more repetitions in our experiments.

A major internal threat to the validity of our work comes from the manual analysis of collected traces.
We need to manually determine whether the \expName{} regions reported by \ourtool{} indeed reveal any tool issue.
Consequently, related evaluation results can be influenced by subjective judgments.
However, it should be noted that any work involving manual judgments in the evaluation is vulnerable to this threat.

\section{Related work}
\noindent \textbf{Automated UI testing for Android.}
There have been various tools over years of development.
The earliest efforts include Monkey~\cite{AndroidMonkey}, a randomized tool that does not consider app UIs or coverage information.
The superior efficiency brings strong competitiveness to the simple tool.
Subsequent efforts result in tools mainly driven by randomness/evolution~\cite{Mao:2016,Machiry:2013,Ye:2013:DFA:2536853.2536881},
UI modeling~\cite{su2017guided,DroidBot,Gu:2019:PGT:3339505.3339542,Hao:2014,Choi:2013},
and systematic exploration~\cite{Azim:2013,EvoDroid,Anand:2012}.
Recent work~\cite{10.1145/3377811.3380402} proposes time-travel testing (referred to as TTT) to help Android UI testing escape from ineffective AUT states including loops and dead ends.
TTT uses checkpoint and restore---checkpointing progressive states and restoring those states after loops and dead ends are detected.

\ourtool{} is different from TTT in terms of design goals.
First, TTT aims to recover from ineffective exploration, while \ourtool{} mainly focuses on prevention.
Second, \expNames{} in \ourtool{} are more general than TTT's lack-of-progress definitions.
One example is logging out, where tools assisted by \ourtool{} can still explore a fraction of app functionalities, such as registering and resetting passwords.
Loops and dead ends are not necessarily present when exploring an app with only a few functionalities.
Third, \ourtool{} aims to enhance existing testing tools without the need to understand their internal design or implementation, instead of building a new testing tool that excels at all apps.

The design of \ourtool{} brings a few advantages.
First, as acknowledged by TTT~\cite{10.1145/3377811.3380402}, the state recovery may lead to inconsistent app states when testing apps with external state dependencies that are maintained at the server side. 
Note that controlling server-side states is challenging, e.g., many industrial apps use external services that the apps have no control of.
\ourtool{}'s preventive strategy avoids this limitation.
Second, \ourtool{}'s preventive strategy does not incur overhead for lack-of-progress detection or state recovery in guided runs.
This strategy is specifically useful when a tool repeatedly gets into \expNames{}.
Third, \ourtool{} does not require additional device support for state recovery (such as RAM data restoring).

\paragraph{\bf Trace Analysis.} Our work is related to log and trace analysis.
Existing work has been focusing on analyzing the logs generated by program code through techniques including anomaly detection~\cite{10.1145/3338906.3338931,8809512,10.1145/3133956.3134015,10.1109/ICSE-NIER.2019.00026},
cause analysis~\cite{10.1145/3341301.3359650,10.5555/2486788.2486842,10.1109/ICSE-Companion.2019.00055,222595,10.1145/3338906.3338961},
failure reproduction~\cite{10.1145/3132747.3132768}, and performance-issue detection~\cite{10.5555/2685048.2685099,10.5555/3026877.3026924}.
Our work focuses on UI traces with the goal of understanding UI \expNames{}, which are different from logs produced by program code.

\paragraph{\bf Parallel Testing.} Our work is also related to parallel testing in the sense of producing multiple variants of the target program (i.e., customizing the AUT by manipulating the UI entries) for the testing tool to work on.
Related work on parallel testing includes parallelizing mutation testing~\cite{doi:10.1002/stvr.1471}, symbolic execution~\cite{10.1145/1966445.1966463,10.1145/1831708.1831732}, and the debugging process~\cite{10.1145/1273463.1273468}. 

\section{Conclusion}
We have exploited the opportunities of improving Android UI testing via automatically identifying and addressing \expNames{}.
Specifically, we have presented \ourtool{}, a general approach and supporting system for effectively identifying and addressing \expNames{}.
We have designed specialized algorithms to support \ourtool{}'s concepts.
Our evaluation results have shown that \ourtool{} identifies \expNames{} that cost up to \maxRegionLength{} of testing time budget, revealing various issues hindering testing efficacy. 
By trying to automatically fix the discovered issues, \ourtool{} helps the Android UI testing tools under evaluation with achieving up to \cumCovImproveMax{} higher code coverage relatively and triggering up to \allCrashImproveMax{} distinct crashes.

\section*{Acknowledgments}
This work was partially supported by 3M Foundation Fellowship, UT Dallas startup funding \#37030034, NSF grants (CNS-1564274, SHF-1816615, CNS-1956007, and CCF-2029049), a Facebook Distributed Systems Research award, Microsoft Azure credits, and Google Cloud credits.

\balance
\bibliographystyle{ACM-Reference-Format}
\bibliography{wenyu}


\begin{thebibliography}{54}


\ifx \showCODEN    \undefined \def \showCODEN     #1{\unskip}     \fi
\ifx \showDOI      \undefined \def \showDOI       #1{#1}\fi
\ifx \showISBNx    \undefined \def \showISBNx     #1{\unskip}     \fi
\ifx \showISBNxiii \undefined \def \showISBNxiii  #1{\unskip}     \fi
\ifx \showISSN     \undefined \def \showISSN      #1{\unskip}     \fi
\ifx \showLCCN     \undefined \def \showLCCN      #1{\unskip}     \fi
\ifx \shownote     \undefined \def \shownote      #1{#1}          \fi
\ifx \showarticletitle \undefined \def \showarticletitle #1{#1}   \fi
\ifx \showURL      \undefined \def \showURL       {\relax}        \fi
\providecommand\bibfield[2]{#2}
\providecommand\bibinfo[2]{#2}
\providecommand\natexlab[1]{#1}
\providecommand\showeprint[2][]{arXiv:#2}

\bibitem[\protect\citeauthoryear{Anand, Naik, Harrold, and Yang}{Anand
  et~al\mbox{.}}{2012}]%
        {Anand:2012}
\bibfield{author}{\bibinfo{person}{Saswat Anand}, \bibinfo{person}{Mayur Naik},
  \bibinfo{person}{Mary~Jean Harrold}, {and} \bibinfo{person}{Hongseok Yang}.}
  \bibinfo{year}{2012}\natexlab{}.
\newblock \showarticletitle{{Automated Concolic Testing of Smartphone Apps}}.
  In \bibinfo{booktitle}{\emph{FSE}}.
\newblock


\bibitem[\protect\citeauthoryear{Azim and Neamtiu}{Azim and Neamtiu}{2013}]%
        {Azim:2013}
\bibfield{author}{\bibinfo{person}{Tanzirul Azim} {and} \bibinfo{person}{Iulian
  Neamtiu}.} \bibinfo{year}{2013}\natexlab{}.
\newblock \showarticletitle{{Targeted and Depth-first Exploration for
  Systematic Testing of Android Apps}}. In \bibinfo{booktitle}{\emph{OOPSLA}}.
\newblock


\bibitem[\protect\citeauthoryear{Baek and Bae}{Baek and Bae}{2016}]%
        {10.1145/2970276.2970313}
\bibfield{author}{\bibinfo{person}{Young-Min Baek} {and}
  \bibinfo{person}{Doo-Hwan Bae}.} \bibinfo{year}{2016}\natexlab{}.
\newblock \showarticletitle{{Automated Model-Based Android GUI Testing Using
  Multi-Level GUI Comparison Criteria}}. In \bibinfo{booktitle}{\emph{ASE}}.
\newblock


\bibitem[\protect\citeauthoryear{Brooks}{Brooks}{1995}]%
        {brooks95:mythical}
\bibfield{author}{\bibinfo{person}{Frederick~P. Brooks}.}
  \bibinfo{year}{1995}\natexlab{}.
\newblock \bibinfo{booktitle}{\emph{{The Mythical Man-Month (Anniversary
  Ed.)}}}.
\newblock \bibinfo{publisher}{Addison-Wesley Longman Publishing Co., Inc.}
\newblock


\bibitem[\protect\citeauthoryear{Bucur, Ureche, Zamfir, and Candea}{Bucur
  et~al\mbox{.}}{2011}]%
        {10.1145/1966445.1966463}
\bibfield{author}{\bibinfo{person}{Stefan Bucur}, \bibinfo{person}{Vlad
  Ureche}, \bibinfo{person}{Cristian Zamfir}, {and} \bibinfo{person}{George
  Candea}.} \bibinfo{year}{2011}\natexlab{}.
\newblock \showarticletitle{{Parallel Symbolic Execution for Automated
  Real-World Software Testing}}. In \bibinfo{booktitle}{\emph{EuroSys}}.
\newblock


\bibitem[\protect\citeauthoryear{Chen}{Chen}{2019}]%
        {10.1109/ICSE-Companion.2019.00055}
\bibfield{author}{\bibinfo{person}{An~Ran Chen}.}
  \bibinfo{year}{2019}\natexlab{}.
\newblock \showarticletitle{{An Empirical Study on Leveraging Logs for
  Debugging Production Failures}}. In \bibinfo{booktitle}{\emph{ICSE}}.
\newblock


\bibitem[\protect\citeauthoryear{Choi, Necula, and Sen}{Choi
  et~al\mbox{.}}{2013}]%
        {Choi:2013}
\bibfield{author}{\bibinfo{person}{Wontae Choi}, \bibinfo{person}{George
  Necula}, {and} \bibinfo{person}{Koushik Sen}.}
  \bibinfo{year}{2013}\natexlab{}.
\newblock \showarticletitle{{Guided GUI Testing of Android Apps with Minimal
  Restart and Approximate Learning}}. In \bibinfo{booktitle}{\emph{OOPSLA}}.
\newblock


\bibitem[\protect\citeauthoryear{Choudhary, Gorla, and Orso}{Choudhary
  et~al\mbox{.}}{2015}]%
        {ChoudharyGO15}
\bibfield{author}{\bibinfo{person}{Shauvik~Roy Choudhary},
  \bibinfo{person}{Alessandra Gorla}, {and} \bibinfo{person}{Alessandro Orso}.}
  \bibinfo{year}{2015}\natexlab{}.
\newblock \showarticletitle{{Automated Test Input Generation for Android: Are
  We There Yet?}}. In \bibinfo{booktitle}{\emph{ASE}}.
\newblock


\bibitem[\protect\citeauthoryear{Cui, Ge, Kasikci, Niu, Sharma, Wang, and
  Yun}{Cui et~al\mbox{.}}{2018}]%
        {222595}
\bibfield{author}{\bibinfo{person}{Weidong Cui}, \bibinfo{person}{Xinyang Ge},
  \bibinfo{person}{Baris Kasikci}, \bibinfo{person}{Ben Niu},
  \bibinfo{person}{Upamanyu Sharma}, \bibinfo{person}{Ruoyu Wang}, {and}
  \bibinfo{person}{Insu Yun}.} \bibinfo{year}{2018}\natexlab{}.
\newblock \showarticletitle{{REPT: Reverse Debugging of Failures in Deployed
  Software}}. In \bibinfo{booktitle}{\emph{OSDI}}.
\newblock


\bibitem[\protect\citeauthoryear{Dong, B\"{o}hme, Cojocaru, and
  Roychoudhury}{Dong et~al\mbox{.}}{2020}]%
        {10.1145/3377811.3380402}
\bibfield{author}{\bibinfo{person}{Zhen Dong}, \bibinfo{person}{Marcel
  B\"{o}hme}, \bibinfo{person}{Lucia Cojocaru}, {and} \bibinfo{person}{Abhik
  Roychoudhury}.} \bibinfo{year}{2020}\natexlab{}.
\newblock \showarticletitle{{Time-Travel Testing of Android Apps}}. In
  \bibinfo{booktitle}{\emph{ICSE}}.
\newblock


\bibitem[\protect\citeauthoryear{Du, Li, Zheng, and Srikumar}{Du
  et~al\mbox{.}}{2017}]%
        {10.1145/3133956.3134015}
\bibfield{author}{\bibinfo{person}{Min Du}, \bibinfo{person}{Feifei Li},
  \bibinfo{person}{Guineng Zheng}, {and} \bibinfo{person}{Vivek Srikumar}.}
  \bibinfo{year}{2017}\natexlab{}.
\newblock \showarticletitle{{DeepLog: Anomaly Detection and Diagnosis from
  System Logs through Deep Learning}}. In \bibinfo{booktitle}{\emph{CCS}}.
\newblock


\bibitem[\protect\citeauthoryear{Google}{Google}{2021a}]%
        {AndroidMonkey}
\bibfield{author}{\bibinfo{person}{Google}.} \bibinfo{year}{2021}\natexlab{a}.
\newblock \bibinfo{title}{{Android Monkey}}.
\newblock
\newblock
\urldef\tempurl%
\url{https://developer.android.com/studio/test/monkey}
\showURL{%
\tempurl}


\bibitem[\protect\citeauthoryear{Google}{Google}{2021b}]%
        {AndroidView}
\bibfield{author}{\bibinfo{person}{Google}.} \bibinfo{year}{2021}\natexlab{b}.
\newblock \bibinfo{title}{{Android View}}.
\newblock
\newblock
\urldef\tempurl%
\url{https://developer.android.com/reference/android/view/View}
\showURL{%
\tempurl}


\bibitem[\protect\citeauthoryear{Google}{Google}{2021c}]%
        {AndroidViewGroup}
\bibfield{author}{\bibinfo{person}{Google}.} \bibinfo{year}{2021}\natexlab{c}.
\newblock \bibinfo{title}{{Android ViewGroup}}.
\newblock
\newblock
\urldef\tempurl%
\url{https://developer.android.com/reference/android/view/ViewGroup}
\showURL{%
\tempurl}


\bibitem[\protect\citeauthoryear{Google}{Google}{2021d}]%
        {logcat}
\bibfield{author}{\bibinfo{person}{Google}.} \bibinfo{year}{2021}\natexlab{d}.
\newblock \bibinfo{title}{{Logcat command-line tool}}.
\newblock
\newblock
\urldef\tempurl%
\url{https://developer.android.com/studio/command-line/logcat}
\showURL{%
\tempurl}


\bibitem[\protect\citeauthoryear{Gu, Cao, Liu, Sun, Deng, Ma, and Lü}{Gu
  et~al\mbox{.}}{2017}]%
        {AimDroid}
\bibfield{author}{\bibinfo{person}{Tianxiao Gu}, \bibinfo{person}{Chun Cao},
  \bibinfo{person}{Tianchi Liu}, \bibinfo{person}{Chengnian Sun},
  \bibinfo{person}{Jing Deng}, \bibinfo{person}{Xiaoxing Ma}, {and}
  \bibinfo{person}{Jian Lü}.} \bibinfo{year}{2017}\natexlab{}.
\newblock \showarticletitle{{AimDroid: Activity-Insulated Multi-level Automated
  Testing for Android Applications}}. In \bibinfo{booktitle}{\emph{ICSME}}.
\newblock


\bibitem[\protect\citeauthoryear{Gu, Sun, Ma, Cao, Xu, Yao, Zhang, Lu, and
  Su}{Gu et~al\mbox{.}}{2019}]%
        {Gu:2019:PGT:3339505.3339542}
\bibfield{author}{\bibinfo{person}{Tianxiao Gu}, \bibinfo{person}{Chengnian
  Sun}, \bibinfo{person}{Xiaoxing Ma}, \bibinfo{person}{Chun Cao},
  \bibinfo{person}{Chang Xu}, \bibinfo{person}{Yuan Yao},
  \bibinfo{person}{Qirun Zhang}, \bibinfo{person}{Jian Lu}, {and}
  \bibinfo{person}{Zhendong Su}.} \bibinfo{year}{2019}\natexlab{}.
\newblock \showarticletitle{{Practical GUI Testing of Android Applications via
  Model Abstraction and Refinement}}. In \bibinfo{booktitle}{\emph{ICSE}}.
\newblock


\bibitem[\protect\citeauthoryear{Hao, Liu, Nath, Halfond, and Govindan}{Hao
  et~al\mbox{.}}{2014}]%
        {Hao:2014}
\bibfield{author}{\bibinfo{person}{Shuai Hao}, \bibinfo{person}{Bin Liu},
  \bibinfo{person}{Suman Nath}, \bibinfo{person}{William~G.J. Halfond}, {and}
  \bibinfo{person}{Ramesh Govindan}.} \bibinfo{year}{2014}\natexlab{}.
\newblock \showarticletitle{{PUMA: Programmable UI-automation for Large-scale
  Dynamic Analysis of Mobile Apps}}. In \bibinfo{booktitle}{\emph{MobiSys}}.
\newblock


\bibitem[\protect\citeauthoryear{IDC and Gartner}{IDC and Gartner}{2019}]%
        {android_market_share}
\bibfield{author}{\bibinfo{person}{IDC} {and} \bibinfo{person}{Gartner}.}
  \bibinfo{year}{2019}\natexlab{}.
\newblock \bibinfo{title}{{Share of Android Os of Global Smartphone Shipments
  from 1st Quarter 2011 to 2nd Quarter 2018}}.
\newblock
\newblock
\urldef\tempurl%
\url{https://www.statista.com/statistics/236027/global-smartphone-os-market-share-of-android/}
\showURL{%
\tempurl}


\bibitem[\protect\citeauthoryear{Jones, Bowring, and Harrold}{Jones
  et~al\mbox{.}}{2007}]%
        {10.1145/1273463.1273468}
\bibfield{author}{\bibinfo{person}{James~A. Jones}, \bibinfo{person}{James~F.
  Bowring}, {and} \bibinfo{person}{Mary~Jean Harrold}.}
  \bibinfo{year}{2007}\natexlab{}.
\newblock \showarticletitle{{Debugging in Parallel}}. In
  \bibinfo{booktitle}{\emph{ISSTA}}.
\newblock


\bibitem[\protect\citeauthoryear{Labs}{Labs}{2018}]%
        {ten_best_tools}
\bibfield{author}{\bibinfo{person}{Mobile Labs}.}
  \bibinfo{year}{2018}\natexlab{}.
\newblock \bibinfo{title}{{10 Best Android \& iOS Automation App Testing
  Tools}}.
\newblock
\newblock
\urldef\tempurl%
\url{https://mobilelabsinc.com/blog/top-10-automated-testing-tools-for-mobile-app-testing}
\showURL{%
\tempurl}


\bibitem[\protect\citeauthoryear{Li, Yang, Guo, and Chen}{Li
  et~al\mbox{.}}{2017}]%
        {DroidBot}
\bibfield{author}{\bibinfo{person}{Yuanchun Li}, \bibinfo{person}{Ziyue Yang},
  \bibinfo{person}{Yao Guo}, {and} \bibinfo{person}{Xiangqun Chen}.}
  \bibinfo{year}{2017}\natexlab{}.
\newblock \showarticletitle{{DroidBot: A Lightweight UI-guided Test Input
  Generator for Android}}. In \bibinfo{booktitle}{\emph{ICSE-C}}.
\newblock


\bibitem[\protect\citeauthoryear{Liu, Craft, Situ, Yumer, Mech, and Kumar}{Liu
  et~al\mbox{.}}{2018}]%
        {10.1145/3242587.3242650}
\bibfield{author}{\bibinfo{person}{Thomas~F. Liu}, \bibinfo{person}{Mark
  Craft}, \bibinfo{person}{Jason Situ}, \bibinfo{person}{Ersin Yumer},
  \bibinfo{person}{Radomir Mech}, {and} \bibinfo{person}{Ranjitha Kumar}.}
  \bibinfo{year}{2018}\natexlab{}.
\newblock \showarticletitle{{Learning Design Semantics for Mobile Apps}}. In
  \bibinfo{booktitle}{\emph{UIST}}.
\newblock


\bibitem[\protect\citeauthoryear{Machiry, Tahiliani, and Naik}{Machiry
  et~al\mbox{.}}{2013}]%
        {Machiry:2013}
\bibfield{author}{\bibinfo{person}{Aravind Machiry}, \bibinfo{person}{Rohan
  Tahiliani}, {and} \bibinfo{person}{Mayur Naik}.}
  \bibinfo{year}{2013}\natexlab{}.
\newblock \showarticletitle{{Dynodroid: An Input Generation System for Android
  Apps}}. In \bibinfo{booktitle}{\emph{ESEC/FSE}}.
\newblock


\bibitem[\protect\citeauthoryear{Mahmood, Mirzaei, and Malek}{Mahmood
  et~al\mbox{.}}{2014}]%
        {EvoDroid}
\bibfield{author}{\bibinfo{person}{Riyadh Mahmood}, \bibinfo{person}{Nariman
  Mirzaei}, {and} \bibinfo{person}{Sam Malek}.}
  \bibinfo{year}{2014}\natexlab{}.
\newblock \showarticletitle{{EvoDroid: Segmented Evolutionary Testing of
  Android Apps}}. In \bibinfo{booktitle}{\emph{FSE}}.
\newblock


\bibitem[\protect\citeauthoryear{Mao, Harman, and Jia}{Mao
  et~al\mbox{.}}{2016}]%
        {Mao:2016}
\bibfield{author}{\bibinfo{person}{Ke Mao}, \bibinfo{person}{Mark Harman},
  {and} \bibinfo{person}{Yue Jia}.} \bibinfo{year}{2016}\natexlab{}.
\newblock \showarticletitle{{Sapienz: Multi-objective Automated Testing for
  Android Applications}}. In \bibinfo{booktitle}{\emph{ISSTA}}.
\newblock


\bibitem[\protect\citeauthoryear{Mateo and Usaola}{Mateo and Usaola}{2013}]%
        {doi:10.1002/stvr.1471}
\bibfield{author}{\bibinfo{person}{Pedro~Reales Mateo} {and}
  \bibinfo{person}{Macario~Polo Usaola}.} \bibinfo{year}{2013}\natexlab{}.
\newblock \showarticletitle{{Parallel mutation testing}}.
\newblock \bibinfo{journal}{\emph{Software Testing, Verification and
  Reliability}} (\bibinfo{year}{2013}).
\newblock


\bibitem[\protect\citeauthoryear{Monni and Pezz\`{e}}{Monni and
  Pezz\`{e}}{2019}]%
        {10.1109/ICSE-NIER.2019.00026}
\bibfield{author}{\bibinfo{person}{Cristina Monni} {and} \bibinfo{person}{Mauro
  Pezz\`{e}}.} \bibinfo{year}{2019}\natexlab{}.
\newblock \showarticletitle{{Energy-Based Anomaly Detection a New Perspective
  for Predicting Software Failures}}. In \bibinfo{booktitle}{\emph{ICSE-NIER}}.
\newblock


\bibitem[\protect\citeauthoryear{MoQuality}{MoQuality}{2021}]%
        {How-to-Automate-Mobile-App-Testing}
\bibfield{author}{\bibinfo{person}{MoQuality}.}
  \bibinfo{year}{2021}\natexlab{}.
\newblock \bibinfo{title}{{How to Automate Mobile App Testing}}.
\newblock
\newblock
\urldef\tempurl%
\url{https://www.moquality.com/blog/How-to-Automate-Mobile-App-Testing}
\showURL{%
\tempurl}


\bibitem[\protect\citeauthoryear{Packevi{\v{c}}ius, Barisas, U{\v{s}}aniov,
  Guogis, and Barei{\v{s}}a}{Packevi{\v{c}}ius et~al\mbox{.}}{2018}]%
        {Packevicius2018TextSA}
\bibfield{author}{\bibinfo{person}{{\v{S}}ar{\={u}}nas Packevi{\v{c}}ius},
  \bibinfo{person}{Dominykas Barisas}, \bibinfo{person}{Andrej U{\v{s}}aniov},
  \bibinfo{person}{Evaldas Guogis}, {and} \bibinfo{person}{Eduardas
  Barei{\v{s}}a}.} \bibinfo{year}{2018}\natexlab{}.
\newblock \showarticletitle{Text Semantics and Layout Defects Detection in
  Android Apps Using Dynamic Execution and Screenshot Analysis}. In
  \bibinfo{booktitle}{\emph{ICIST}}.
\newblock


\bibitem[\protect\citeauthoryear{Schrijver}{Schrijver}{1986}]%
        {linear_programming}
\bibfield{author}{\bibinfo{person}{Alexander Schrijver}.}
  \bibinfo{year}{1986}\natexlab{}.
\newblock \bibinfo{booktitle}{\emph{{Theory of Linear and Integer
  Programming}}}.
\newblock \bibinfo{publisher}{John Wiley \& Sons, Inc.}
\newblock


\bibitem[\protect\citeauthoryear{Shang, Jiang, Hemmati, Adams, Hassan, and
  Martin}{Shang et~al\mbox{.}}{2013}]%
        {10.5555/2486788.2486842}
\bibfield{author}{\bibinfo{person}{Weiyi Shang}, \bibinfo{person}{Zhen~Ming
  Jiang}, \bibinfo{person}{Hadi Hemmati}, \bibinfo{person}{Bram Adams},
  \bibinfo{person}{Ahmed~E. Hassan}, {and} \bibinfo{person}{Patrick Martin}.}
  \bibinfo{year}{2013}\natexlab{}.
\newblock \showarticletitle{Assisting Developers of Big Data Analytics
  Applications When Deploying on Hadoop Clouds}. In
  \bibinfo{booktitle}{\emph{ICSE}}.
\newblock


\bibitem[\protect\citeauthoryear{SmartBear}{SmartBear}{2021}]%
        {mastering_the_art_of_mobile_testing}
\bibfield{author}{\bibinfo{person}{SmartBear}.}
  \bibinfo{year}{2021}\natexlab{}.
\newblock \bibinfo{title}{{Mastering the Art of Mobile Testing}}.
\newblock
\newblock
\urldef\tempurl%
\url{https://smartbear.com/resources/ebooks/mastering-the-art-of-mobile-testing/}
\showURL{%
\tempurl}


\bibitem[\protect\citeauthoryear{Staats and Pundefinedsundefinedreanu}{Staats
  and Pundefinedsundefinedreanu}{2010}]%
        {10.1145/1831708.1831732}
\bibfield{author}{\bibinfo{person}{Matt Staats} {and} \bibinfo{person}{Corina
  Pundefinedsundefinedreanu}.} \bibinfo{year}{2010}\natexlab{}.
\newblock \showarticletitle{{Parallel Symbolic Execution for Structural Test
  Generation}}. In \bibinfo{booktitle}{\emph{ISSTA}}.
\newblock


\bibitem[\protect\citeauthoryear{Su, Meng, Chen, Wu, Yang, Yao, Pu, Liu, and
  Su}{Su et~al\mbox{.}}{2017}]%
        {su2017guided}
\bibfield{author}{\bibinfo{person}{Ting Su}, \bibinfo{person}{Guozhu Meng},
  \bibinfo{person}{Yuting Chen}, \bibinfo{person}{Ke Wu},
  \bibinfo{person}{Weiming Yang}, \bibinfo{person}{Yao Yao},
  \bibinfo{person}{Geguang Pu}, \bibinfo{person}{Yang Liu}, {and}
  \bibinfo{person}{Zhendong Su}.} \bibinfo{year}{2017}\natexlab{}.
\newblock \showarticletitle{{Guided, Stochastic Model-based GUI Testing of
  Android Apps}}. In \bibinfo{booktitle}{\emph{ESEC/FSE}}.
\newblock


\bibitem[\protect\citeauthoryear{Testsigma}{Testsigma}{2021}]%
        {how_to_perform_mobile_automation_testing}
\bibfield{author}{\bibinfo{person}{Testsigma}.}
  \bibinfo{year}{2021}\natexlab{}.
\newblock \bibinfo{title}{{How to perform Mobile Automation Testing of the
  UI?}}
\newblock
\newblock
\urldef\tempurl%
\url{https://testsigma.com/blog/how-to-perform-mobile-automation-testing-of-the-ui/}
\showURL{%
\tempurl}


\bibitem[\protect\citeauthoryear{{VET Artifacts}}{{VET Artifacts}}{2021}]%
        {vet-github}
\bibfield{author}{\bibinfo{person}{{VET Artifacts}}.}
  \bibinfo{year}{2021}\natexlab{}.
\newblock
\newblock
\newblock
\shownote{\url{https://github.com/VET-UI-Testing/main}.}


\bibitem[\protect\citeauthoryear{Wang, Lam, and Xie}{Wang
  et~al\mbox{.}}{2021}]%
        {toller-paper}
\bibfield{author}{\bibinfo{person}{Wenyu Wang}, \bibinfo{person}{Wing Lam},
  {and} \bibinfo{person}{Tao Xie}.} \bibinfo{year}{2021}\natexlab{}.
\newblock \showarticletitle{{An Infrastructure Approach to Improving
  Effectiveness of Android UI Testing Tools}}. In
  \bibinfo{booktitle}{\emph{ISSTA}}.
\newblock


\bibitem[\protect\citeauthoryear{Wang, Li, Yang, Cao, Zhang, Deng, and
  Xie}{Wang et~al\mbox{.}}{2018}]%
        {Wang:2018:ESA:3238147.3240465}
\bibfield{author}{\bibinfo{person}{Wenyu Wang}, \bibinfo{person}{Dengfeng Li},
  \bibinfo{person}{Wei Yang}, \bibinfo{person}{Yurui Cao},
  \bibinfo{person}{Zhenwen Zhang}, \bibinfo{person}{Yuetang Deng}, {and}
  \bibinfo{person}{Tao Xie}.} \bibinfo{year}{2018}\natexlab{}.
\newblock \showarticletitle{{An Empirical Study of Android Test Generation
  Tools in Industrial Cases}}. In \bibinfo{booktitle}{\emph{ASE}}.
\newblock


\bibitem[\protect\citeauthoryear{Wolsey and Nemhauser}{Wolsey and
  Nemhauser}{1999}]%
        {wolsey1999integer}
\bibfield{author}{\bibinfo{person}{Laurence~A Wolsey} {and}
  \bibinfo{person}{George~L Nemhauser}.} \bibinfo{year}{1999}\natexlab{}.
\newblock \bibinfo{booktitle}{\emph{{Integer and combinatorial optimization}}}.
\newblock \bibinfo{publisher}{John Wiley \& Sons}.
\newblock


\bibitem[\protect\citeauthoryear{Xi, Yang, Xiao, Yao, Xiong, Xu, Wang, Gao,
  Liu, Xu, and Lu}{Xi et~al\mbox{.}}{2019}]%
        {10.1145/3319535.3363193}
\bibfield{author}{\bibinfo{person}{Shengqu Xi}, \bibinfo{person}{Shao Yang},
  \bibinfo{person}{Xusheng Xiao}, \bibinfo{person}{Yuan Yao},
  \bibinfo{person}{Yayuan Xiong}, \bibinfo{person}{Fengyuan Xu},
  \bibinfo{person}{Haoyu Wang}, \bibinfo{person}{Peng Gao},
  \bibinfo{person}{Zhuotao Liu}, \bibinfo{person}{Feng Xu}, {and}
  \bibinfo{person}{Jian Lu}.} \bibinfo{year}{2019}\natexlab{}.
\newblock \showarticletitle{{DeepIntent: Deep Icon-Behavior Learning for
  Detecting Intention-Behavior Discrepancy in Mobile Apps}}. In
  \bibinfo{booktitle}{\emph{CCS}}.
\newblock


\bibitem[\protect\citeauthoryear{Xiao, Wang, Cao, Wang, and Gao}{Xiao
  et~al\mbox{.}}{2019}]%
        {10.1109/ICSE.2019.00041}
\bibfield{author}{\bibinfo{person}{Xusheng Xiao}, \bibinfo{person}{Xiaoyin
  Wang}, \bibinfo{person}{Zhihao Cao}, \bibinfo{person}{Hanlin Wang}, {and}
  \bibinfo{person}{Peng Gao}.} \bibinfo{year}{2019}\natexlab{}.
\newblock \showarticletitle{IconIntent: Automatic Identification of Sensitive
  UI Widgets Based on Icon Classification for Android Apps}. In
  \bibinfo{booktitle}{\emph{ICSE}}.
\newblock


\bibitem[\protect\citeauthoryear{Yang, Prasad, and Xie}{Yang
  et~al\mbox{.}}{2013}]%
        {yang13:grey}
\bibfield{author}{\bibinfo{person}{Wei Yang}, \bibinfo{person}{Mukul~R.
  Prasad}, {and} \bibinfo{person}{Tao Xie}.} \bibinfo{year}{2013}\natexlab{}.
\newblock \showarticletitle{{A Grey-box Approach for Automated GUI-model
  Generation of Mobile Applications}}. In \bibinfo{booktitle}{\emph{FASE}}.
\newblock


\bibitem[\protect\citeauthoryear{Ye, Cheng, Zhang, and Jiang}{Ye
  et~al\mbox{.}}{2013}]%
        {Ye:2013:DFA:2536853.2536881}
\bibfield{author}{\bibinfo{person}{Hui Ye}, \bibinfo{person}{Shaoyin Cheng},
  \bibinfo{person}{Lanbo Zhang}, {and} \bibinfo{person}{Fan Jiang}.}
  \bibinfo{year}{2013}\natexlab{}.
\newblock \showarticletitle{{DroidFuzzer: Fuzzing the Android Apps with
  Intent-Filter Tag}}. In \bibinfo{booktitle}{\emph{MoMM}}.
\newblock


\bibitem[\protect\citeauthoryear{Zeng, Li, Zheng, Xia, Deng, Lam, Yang, and
  Xie}{Zeng et~al\mbox{.}}{2016}]%
        {Zeng:2016}
\bibfield{author}{\bibinfo{person}{Xia Zeng}, \bibinfo{person}{Dengfeng Li},
  \bibinfo{person}{Wujie Zheng}, \bibinfo{person}{Fan Xia},
  \bibinfo{person}{Yuetang Deng}, \bibinfo{person}{Wing Lam},
  \bibinfo{person}{Wei Yang}, {and} \bibinfo{person}{Tao Xie}.}
  \bibinfo{year}{2016}\natexlab{}.
\newblock \showarticletitle{{Automated Test Input Generation for Android: Are
  We Really There Yet in an Industrial Case?}}. In
  \bibinfo{booktitle}{\emph{FSE}}.
\newblock


\bibitem[\protect\citeauthoryear{Zhang and Shasha}{Zhang and Shasha}{1989}]%
        {tree_edit_dist}
\bibfield{author}{\bibinfo{person}{Kaizhong Zhang} {and}
  \bibinfo{person}{Dennis Shasha}.} \bibinfo{year}{1989}\natexlab{}.
\newblock \showarticletitle{{Simple Fast Algorithms for the Editing Distance
  Between Trees and Related Problems}}.
\newblock \bibinfo{journal}{\emph{SIAM J. Comput.}} (\bibinfo{date}{12}
  \bibinfo{year}{1989}).
\newblock


\bibitem[\protect\citeauthoryear{Zhang, Xu, Lin, Qiao, Zhang, Dang, Xie, Yang,
  Cheng, Li, Chen, He, Yao, Lou, Chintalapati, Shen, and Zhang}{Zhang
  et~al\mbox{.}}{2019b}]%
        {10.1145/3338906.3338931}
\bibfield{author}{\bibinfo{person}{Xu Zhang}, \bibinfo{person}{Yong Xu},
  \bibinfo{person}{Qingwei Lin}, \bibinfo{person}{Bo Qiao},
  \bibinfo{person}{Hongyu Zhang}, \bibinfo{person}{Yingnong Dang},
  \bibinfo{person}{Chunyu Xie}, \bibinfo{person}{Xinsheng Yang},
  \bibinfo{person}{Qian Cheng}, \bibinfo{person}{Ze Li},
  \bibinfo{person}{Junjie Chen}, \bibinfo{person}{Xiaoting He},
  \bibinfo{person}{Randolph Yao}, \bibinfo{person}{Jian-Guang Lou},
  \bibinfo{person}{Murali Chintalapati}, \bibinfo{person}{Furao Shen}, {and}
  \bibinfo{person}{Dongmei Zhang}.} \bibinfo{year}{2019}\natexlab{b}.
\newblock \showarticletitle{{Robust Log-Based Anomaly Detection on Unstable Log
  Data}}. In \bibinfo{booktitle}{\emph{ESEC/FSE}}.
\newblock


\bibitem[\protect\citeauthoryear{Zhang, Makarov, Ren, Lion, and Yuan}{Zhang
  et~al\mbox{.}}{2017}]%
        {10.1145/3132747.3132768}
\bibfield{author}{\bibinfo{person}{Yongle Zhang}, \bibinfo{person}{Serguei
  Makarov}, \bibinfo{person}{Xiang Ren}, \bibinfo{person}{David Lion}, {and}
  \bibinfo{person}{Ding Yuan}.} \bibinfo{year}{2017}\natexlab{}.
\newblock \showarticletitle{{Pensieve: Non-Intrusive Failure Reproduction for
  Distributed Systems Using the Event Chaining Approach}}. In
  \bibinfo{booktitle}{\emph{SOSP}}.
\newblock


\bibitem[\protect\citeauthoryear{Zhang, Rodrigues, Luo, Stumm, and Yuan}{Zhang
  et~al\mbox{.}}{2019a}]%
        {10.1145/3341301.3359650}
\bibfield{author}{\bibinfo{person}{Yongle Zhang}, \bibinfo{person}{Kirk
  Rodrigues}, \bibinfo{person}{Yu Luo}, \bibinfo{person}{Michael Stumm}, {and}
  \bibinfo{person}{Ding Yuan}.} \bibinfo{year}{2019}\natexlab{a}.
\newblock \showarticletitle{{The Inflection Point Hypothesis: A Principled
  Debugging Approach for Locating the Root Cause of a Failure}}. In
  \bibinfo{booktitle}{\emph{SOSP}}.
\newblock


\bibitem[\protect\citeauthoryear{Zhao, Rodrigues, Luo, Yuan, and Stumm}{Zhao
  et~al\mbox{.}}{2016}]%
        {10.5555/3026877.3026924}
\bibfield{author}{\bibinfo{person}{Xu Zhao}, \bibinfo{person}{Kirk Rodrigues},
  \bibinfo{person}{Yu Luo}, \bibinfo{person}{Ding Yuan}, {and}
  \bibinfo{person}{Michael Stumm}.} \bibinfo{year}{2016}\natexlab{}.
\newblock \showarticletitle{{Non-Intrusive Performance Profiling for Entire
  Software Stacks Based on the Flow Reconstruction Principle}}. In
  \bibinfo{booktitle}{\emph{OSDI}}.
\newblock


\bibitem[\protect\citeauthoryear{Zhao, Zhang, Lion, Ullah, Luo, Yuan, and
  Stumm}{Zhao et~al\mbox{.}}{2014}]%
        {10.5555/2685048.2685099}
\bibfield{author}{\bibinfo{person}{Xu Zhao}, \bibinfo{person}{Yongle Zhang},
  \bibinfo{person}{David Lion}, \bibinfo{person}{Muhammad~Faizan Ullah},
  \bibinfo{person}{Yu Luo}, \bibinfo{person}{Ding Yuan}, {and}
  \bibinfo{person}{Michael Stumm}.} \bibinfo{year}{2014}\natexlab{}.
\newblock \showarticletitle{{Lprof: A Non-Intrusive Request Flow Profiler for
  Distributed Systems}}. In \bibinfo{booktitle}{\emph{OSDI}}.
\newblock


\bibitem[\protect\citeauthoryear{Zhao, Cerf, Birke, Robu, Bouchenak,
  Ben~Mokhtar, and Chen}{Zhao et~al\mbox{.}}{2019}]%
        {8809512}
\bibfield{author}{\bibinfo{person}{Zilong Zhao}, \bibinfo{person}{Sophie Cerf},
  \bibinfo{person}{Robert Birke}, \bibinfo{person}{Bogdan Robu},
  \bibinfo{person}{Sara Bouchenak}, \bibinfo{person}{Sonia Ben~Mokhtar}, {and}
  \bibinfo{person}{Lydia~Y Chen}.} \bibinfo{year}{2019}\natexlab{}.
\newblock \showarticletitle{Robust Anomaly Detection on Unreliable Data}. In
  \bibinfo{booktitle}{\emph{DSN}}.
\newblock


\bibitem[\protect\citeauthoryear{Zheng, Li, Liang, Zeng, Zheng, Deng, Lam,
  Yang, and Xie}{Zheng et~al\mbox{.}}{2017}]%
        {zheng17:automated}
\bibfield{author}{\bibinfo{person}{Haibing Zheng}, \bibinfo{person}{Dengfeng
  Li}, \bibinfo{person}{Beihai Liang}, \bibinfo{person}{Xia Zeng},
  \bibinfo{person}{Wujie Zheng}, \bibinfo{person}{Yuetang Deng},
  \bibinfo{person}{Wing Lam}, \bibinfo{person}{Wei Yang}, {and}
  \bibinfo{person}{Tao Xie}.} \bibinfo{year}{2017}\natexlab{}.
\newblock \showarticletitle{{Automated Test Input Generation for Android:
  Towards Getting There in an Industrial Case}}. In
  \bibinfo{booktitle}{\emph{ICSE-SEIP}}.
\newblock


\bibitem[\protect\citeauthoryear{Zhou, Peng, Xie, Sun, Ji, Liu, Xiang, and
  He}{Zhou et~al\mbox{.}}{2019}]%
        {10.1145/3338906.3338961}
\bibfield{author}{\bibinfo{person}{Xiang Zhou}, \bibinfo{person}{Xin Peng},
  \bibinfo{person}{Tao Xie}, \bibinfo{person}{Jun Sun}, \bibinfo{person}{Chao
  Ji}, \bibinfo{person}{Dewei Liu}, \bibinfo{person}{Qilin Xiang}, {and}
  \bibinfo{person}{Chuan He}.} \bibinfo{year}{2019}\natexlab{}.
\newblock \showarticletitle{{Latent Error Prediction and Fault Localization for
  Microservice Applications by Learning from System Trace Logs}}. In
  \bibinfo{booktitle}{\emph{ESEC/FSE}}.
\newblock


\end{thebibliography}

\end{document}